\documentclass[sigconf]{acmart}

\usepackage{booktabs} 
\usepackage{comment}
\usepackage{balance}
\usepackage{graphics}

\copyrightyear{2020}
\acmYear{2020}
\setcopyright{acmlicensed}\acmConference[CPSS '20]{Proceedings of the 6th ACM Cyber-Physical System Security Workshop}{October 6, 2020}{Taipei, Taiwan}
\acmBooktitle{Proceedings of the 6th ACM Cyber-Physical System Security Workshop (CPSS '20), October 6, 2020, Taipei, Taiwan}
\acmPrice{15.00}
\acmDOI{10.1145/3384941.3409589}
\acmISBN{978-1-4503-7608-2/20/10}

\newcommand{\substepseparator}{\hspace{1cm}}

\begin{document}
\fancyhead{}

\title{Towards Systematically Deriving Defence Mechanisms from Functional Requirements of Cyber-Physical Systems}

\author{Cheah Huei Yoong}
\affiliation{\institution{Singapore University of Technology and Design}}
\email{cheahhuei_yoong@sutd.edu.sg}

\author{Venkata Reddy Palleti}
\affiliation{\institution{Indian Institute of Petroleum and Energy}}
\email{venkat_palleti.che@iipe.ac.in}

\author{Arlindo Silva}
\orcid{0000-0001-5120-3914}
\affiliation{\institution{Singapore University of Technology and Design}}
\email{arlindo_silva@sutd.edu.sg}

\author{Christopher M. Poskitt}
\orcid{0000-0002-9376-2471}
\affiliation{\institution{Singapore Management University}}
\email{cposkitt@smu.edu.sg}

\begin{abstract}
    The threats faced by cyber-physical systems~(CPSs) in critical infrastructure have motivated the development of different attack detection mechanisms, such as those that monitor for violations of \emph{invariants}, i.e.~properties that always hold in normal operation. Given the complexity of CPSs, several existing approaches focus on deriving invariants automatically from data logs, but these can miss possible system behaviours if they are not represented in that data. Furthermore, resolving any design flaws identified in this process is costly, as the CPS is already built. In this position paper, we propose a systematic method for deriving invariants \emph{before} a CPS is built by analysing its \emph{functional requirements}. Our method, inspired by the \emph{axiomatic design} methodology for systems, iteratively analyses dependencies in the design to construct equations and process graphs that model the invariant relationships between CPS components. As a preliminary study, we applied it to the design of a water treatment plant testbed, implementing checkers for two invariants by using decision trees, and finding that they could detect some examples of attacks on the testbed with high accuracy and without false positives. Finally, we explore how developing our method further could lead to more robust CPSs and reduced costs by identifying design weaknesses before systems are implemented.

\end{abstract}

%
%



\maketitle

\section{Introduction}

Cyber-physical systems (CPSs), in which software components and physical processes are tightly integrated, are prevalent in the automation of critical infrastructure, e.g.~as the industrial control systems of power grids and water purification plants. The potentially serious consequences of such systems being compromised~\cite{Hassanzadeh-et_al19a,Leyden16a,ICS-Cert-Alert16a} has motivated the development of different countermeasures for attack detection and prevention, including techniques based on anomaly detection~\cite{Cheng-Tian-Yao17a,Goh_et-al17a,Harada-et_al17a,Inoue-et_al17a,Pasqualetti-Dorfler-Bullo11a,Aggarwal-et_al18a,Aoudi-et_al18a,He-et_al19a,Kravchik-Shabtai18a,Lin-et_al18a,Narayanan-Bobba18a,Schneider-Boettinger18a,Carrasco-Wu19a,Kim-Yun-Kim19a,Adepu-et_al20a,Das-Adepu-Zhou20a}, fingerprinting~\cite{Ahmed-et_al18a,Ahmed-et_al18b,Formby-et_al16a,Gu-et_al18a,Kneib-Huth18a,Yang-et_al20a}, and fuzzing~\cite{Chen-Poskitt-et_al19a,Chen-Xuan-Poskitt-et_al20a,Wijaya-Aniche-Mathur20a}.

Another popular approach is to monitor \emph{invariants} of a CPS~\cite{Giraldo-et_al18a}, i.e.~properties that \emph{always} hold under normal operating conditions, and the violation of which might suggest the presence of an attacker in the system. Invariants are typically relations over the sensor readings and actuator states of a system, a simple example being that ``if the tank level is above $x$, then pump $p$ should be ON''. Given the complexity of CPSs in general, several approaches (e.g.~\cite{Chen-Poskitt-Sun16a,Chen-Poskitt-Sun18a,Feng-et_al19a}) aim to derive such invariants automatically from sources of data, for instance, the time series of sensor readings and actuator states logged by a supervisory control and data acquisition system (SCADA). There is a risk, however, that viable system behaviours are missed if they are not represented in that data (e.g.~rarely occurring), and addressing any design flaws identified is costly as the CPS is already built. Invariants can be derived manually by system engineers~\cite{Cardenas-et_al11a,Adepu-Mathur16a,Adepu-Mathur16b,Adepu-Mathur18b,Choi-et_al18a}, but if done so in an ad hoc manner, may also lead to properties being missed.

In this position paper, we propose a systematic method to support the derivation of invariants \emph{before} a CPS is built by analysing its \emph{functional requirements}. Our goal is to further integrate security concerns in the design stage, ensuring that invariants can be traced from requirements through to implemented defence mechanisms, and to potentially save costs by identifying weak points before the design is executed. Our approach, inspired by \emph{axiomatic design}~\cite{Suh01a}---a design science methodology for systems---iteratively analyses dependencies in the CPS to construct equations and process graphs that model the required invariant relationships between sensors and actuators. Checkers for these invariants can then be implemented using techniques such as decision trees, and when deployed with the built CPS, can monitor its data for any violations of the properties.

To evaluate the viability of our proposals, we applied our method to the design of Secure Water Treatment~(SWaT)~\cite{SWaT-Reference,Mathur-Tippenhauer16a}, a scaled-down version of a real-world water purification plant. SWaT is a complex multi-stage CPS involving physical and chemical processes such as ultrafiltration, de-chlorination, and reverse osmosis. We used our method to decompose eight high-level functional requirements into 44 concrete ones, tracing dependencies down to specific sensors and actuators. We illustrate how this information can be used to construct invariants, represented either as equations or process graphs. Finally, as a preliminary study, we implemented checkers for two of the invariants by training supervised machine learning~(ML) models (i.e.~decision trees) on equation inputs/outputs, finding that they could detect relevant attacks in both a dataset \emph{and} on the real testbed with high levels of accuracy and without false positives.

\substepseparator

\noindent\textbf{Summary of Contributions.} We propose a method for systematically deriving invariants from the functional requirements of a CPS. Through a preliminary study on a real-world critical infrastructure testbed, we demonstrate the feasibility of constructing effective invariant-based checkers for CPSs that can be traced all the way back to the system's requirements. Furthermore, our study demonstrates the potential of a cross-disciplinary approach that utilises methodologies from design science (i.e.~axiomatic design) to structure and analyse the dependencies present in critical infrastructure.

This work extends the initial ideas of Palleti et al.~\cite{Palleti-Joseph-Silva18a}, who explored how axiomatic design could help analyse the requirements of a water distribution network. We go further by applying it to a water purification plant and deriving invariant-based checkers that can be used for its defence.

\substepseparator

\noindent\textbf{Organisation.} The rest of the paper is organised as follows. Section~\ref{sec:background} briefly describes the SWaT architecture, as well as the axiomatic design theory our method is inspired by. Next, our systematic design framework is presented in Section~\ref{sec:main}. Section~\ref{sec:study} discusses the effectiveness of our invariant checkers in a preliminary study. Finally, Section~\ref{sec:conclusion} concludes this paper.

\section{Background}\label{sec:background}

In this section, we present an overview of the water purification plant that forms the case study of this position paper. Following this, we discuss the axiomatic design theory that our proposed approach is based on.

\subsection{SWaT Testbed}

\begin{figure*}[!t]
  \centering
  \includegraphics[trim=0 1 0 0,clip,width = \textwidth ]{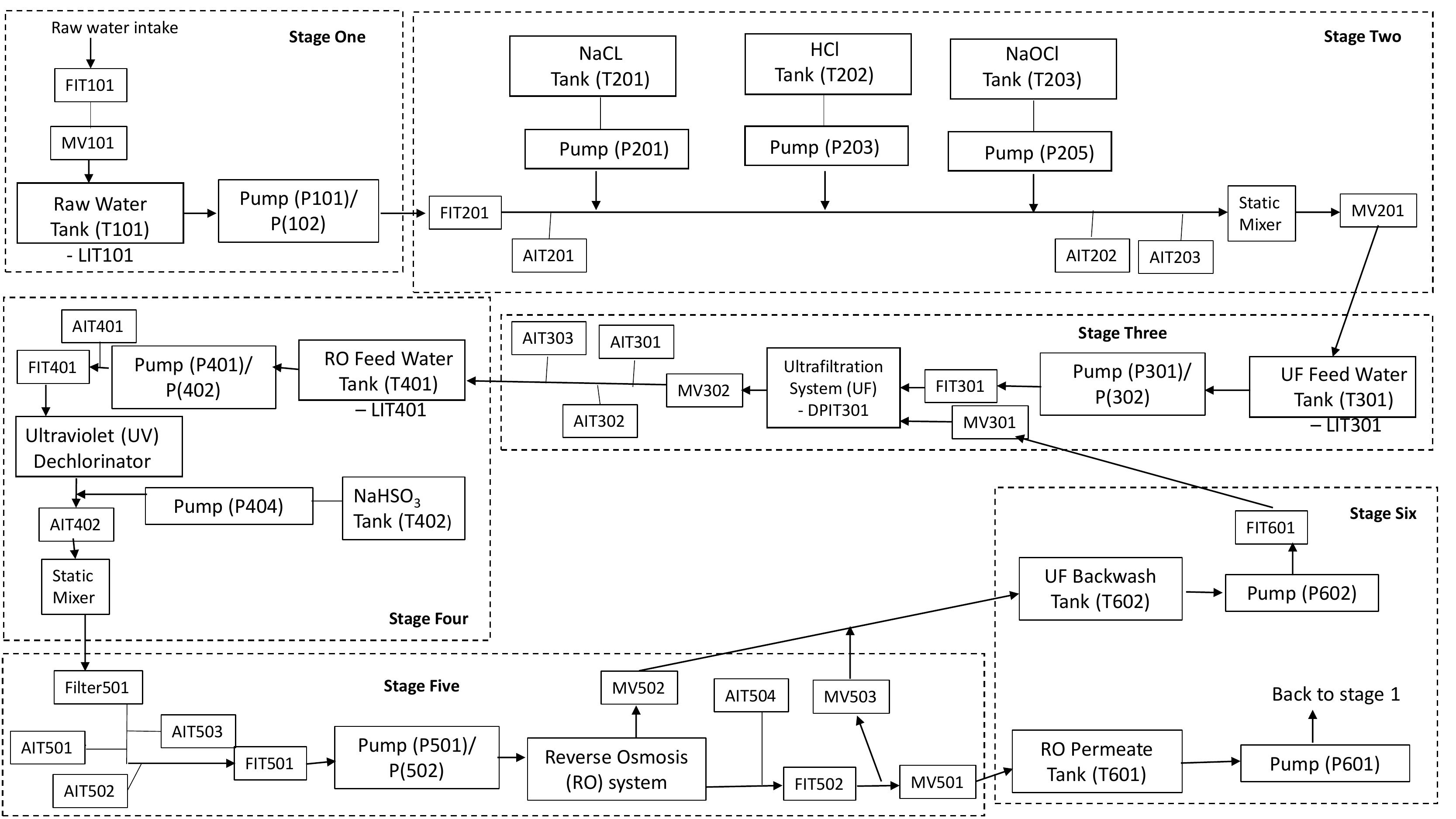}
  \caption{The sub-processes of SWaT}
  \label{fig:swat_processes}
\end{figure*}

The Secure Water Treatment (SWaT) testbed~\cite{SWaT-Reference,Mathur-Tippenhauer16a} is a scaled-down version of a modern water purification plant, intended for supporting research into cyber-security solutions for critical infrastructure. SWaT is able to produce up to five gallons of safe drinking water per minute across six distinct co-operating stages (Figure~\ref{fig:swat_processes}) involving chemical processes like ultrafiltration, de-chlorination, and reverse osmosis. Each stage is controlled by a Programmable Logic Controller~(PLC), which communicates with sensors and actuators through a field-bus network, and with each other through a 24-port Ethernet switch. A SCADA workstation connects a human-machine interface to all of the PLCs, facilitating monitoring and control of the plant by human operators. The physical state of SWaT, as observed by the sensors, is recorded by a historian server at pre-specified intervals. A SWaT dataset is available, consisting of all the data recorded by this server over a period of several days, including a few during which the testbed was subjected to attacks~\cite{CPS-Datasets,Goh-et_al16a}.

An overview of the six sub-processes of SWaT is given in Figure~\ref{fig:swat_processes}. A number of the testbed's 68 sensors and actuators are depicted, with sensors including Flow Indicator Transmitters (FITs), Analyzer Indicator Transmitters (AITs), and Level Indicator Transmitters (LITs). Actuators include Motorised Valves (MVs) for controlling the inflow of water into tanks, and Pumps (Ps) for pumping it out.

\emph{Stage One.} At the bottom of raw water tank T-101, a motorised valve (MV-101) is opened to allow raw water to flow in. An electromagnetic flow transmitter (FIT-101) reads the flow rate of this water, and sends it to the PLC. Pump P-101 transfers water from T-101 to the ultrafiltration feed water tank T-301 in stage three, passing through the chemical dosing of stage two. The operation of P-101 is interlocked to the level indicator transmitter (LIT-301) in tank T-301.

\emph{Stage Two.} Chemical dosing is applied in this stage. The chemical properties of the incoming raw water are measured using analyser indicator transmitters AIT-201, AIT-202, and AIT-203. This information is used by the PLC to control pumps P-201, P-202, and P-203, adjusting the dosing and thus the water's chemical properties before it enters stage three.

\emph{Stage Three.} Ultrafiltration (UF) is performed in this stage. Raw water, after being dosed with chemicals in stage two, is fed into an UF unit. The operation of P-301 is interlocked with the level indicator transmitter LIT-401 for the reverse osmosis (RO) feed water tank (T-401) in stage four. Thus, P-301 is stopped when the water level in T-401 is high, and P-301 is turned on and MV-302 opened when the water level in tank T-401 reaches the low marker. Flow meter FIT-301 measures the incoming flow rate to the UF unit. The differential pressure indicator transmitter (DPIT) continuously monitors the difference in inlet pressure and outlet pressure. If the UF membranes are clogged, the DPIT triggers an alarm and a backwash sequence begins in stage six. AIT-301, AIT-302, and AIT-303 measure and transmit (to the PLC) various chemical parameters of the water entering the UF feed water tank T-301.

\emph{Stage Four.} De-chlorination is performed in this stage: any free chlorine in the water coming out of the UF unit is removed using a combination of an ultraviolet de-chlorinator and sodium bisulphate. P-401 is started when T-401 reaches the high marker, moving water through the de-chlorinator unit.   The hardness analyser (AIT-401) monitors and indicates the level of hardness to avoid scaling within the RO system.

\emph{Stage Five.} Reverse osmosis (RO) is applied in stage five. The RO system is designed to provide bulk reduction of inorganic impurities. The RO permeate stream is channeled to the RO permeate tank (T-601) when MV-501 is opened. Before reaching the tank, the RO permeate conductivity analyser (AIT-504) analyses water conductivity, and if above the threshold, water is diverted to a reject tank T-602 by opening valve MV-503.  The rejected water is used to clean the UF membranes in the backwash process.   RO permeate pump P-601 recycles water from T-601 back to T-101. 

\emph{Stage Six.} Finally, stage six consists of a backwash process. UF membranes need cleaning to remove solid particles. This cleaning is achieved through the backwash process, which is programmed to start every 30 minutes. It is also started when the pressure drop across the membrane goes above a pre-set threshold. The rejected RO water from tank T-602 is moved through the UF unit by starting pump P-602.

\subsection{Axiomatic Design Theory}\label{sec:axiomatic_design}
Axiomatic design is a systems design methodology, developed by Nam Pyo Suh~\cite{Suh01a}, that uses matrix methods to systematically analyse the transformation of customer needs into functional requirements, design parameters, and process variables. The objective of the theory is to create a scientific base for the design process by building upon a suite of fundamental theories from logic and rationale thinking. Researchers have applied this theory in areas such as manufacturing \cite{Matt12a,Zhu-et_al08a} and software development \cite{Kandjani-et_al15a,Mohsen-Cekecek00a}.

In axiomatic design, functional requirements~(FRs) express \emph{what} we want to achieve, i.e.~the specific behaviours we want from the design. Design parameters~(DPs) are elements of the physical design that are chosen to \emph{realise} the FRs. Finally, process variables~(PVs) are elements of the process design controlling the DPs (e.g.~continuous or discrete values characterising the process). Matrix methods are used by the designer to map FRs to DPs in the physical domain. For example, suppose that the top-level of a design involved two FRs and two DPs. These can then be related using the following matrix:
\[
  \begin{bmatrix}
    FR_1 \\
    FR_2
  \end{bmatrix}
  =
  \begin{bmatrix}
    b_{11} & b_{12} \\
    b_{21} & b_{22}
  \end{bmatrix}
  \begin{bmatrix}
    DP_1 \\
    DP_2
  \end{bmatrix}
\]
\noindent The square matrix is a binary (or Boolean) matrix, indicating the coupling between FRs and DPs. After identifying the couplings at a high level (e.g.~where one DP might represent \emph{all} pumps), the designer would decompose the FRs and DPs further (e.g.~with one DP representing exactly \emph{one} of the pumps) until achieving a fine-grained set of dependencies in the design. The decomposed matrices can then be subjected to analyses to assess and mitigate the effects of coupling.

In our work (Section~\ref{sec:main}), axiomatic design is \emph{not} used to design a CPS from the ground-up, but rather, its principles are applied to an existing design to support the derivation of invariants based on the dependencies it implies, and to help assess where weak points might exist (e.g.~due to certain couplings). In SWaT, DPs will correspond to components such as tank level sensors and motorised valves, each of which can function within specific values of PVs. This paves the way for a simple and high-level of analysis to establish the relations between DPs that represent normal CPS behaviour. These relations are then taken as our invariants.


\section{Design Framework}\label{sec:main}

\begin{figure*}[!t]
  \includegraphics[trim=0 70 0 0,clip,width = 0.85\textwidth ]{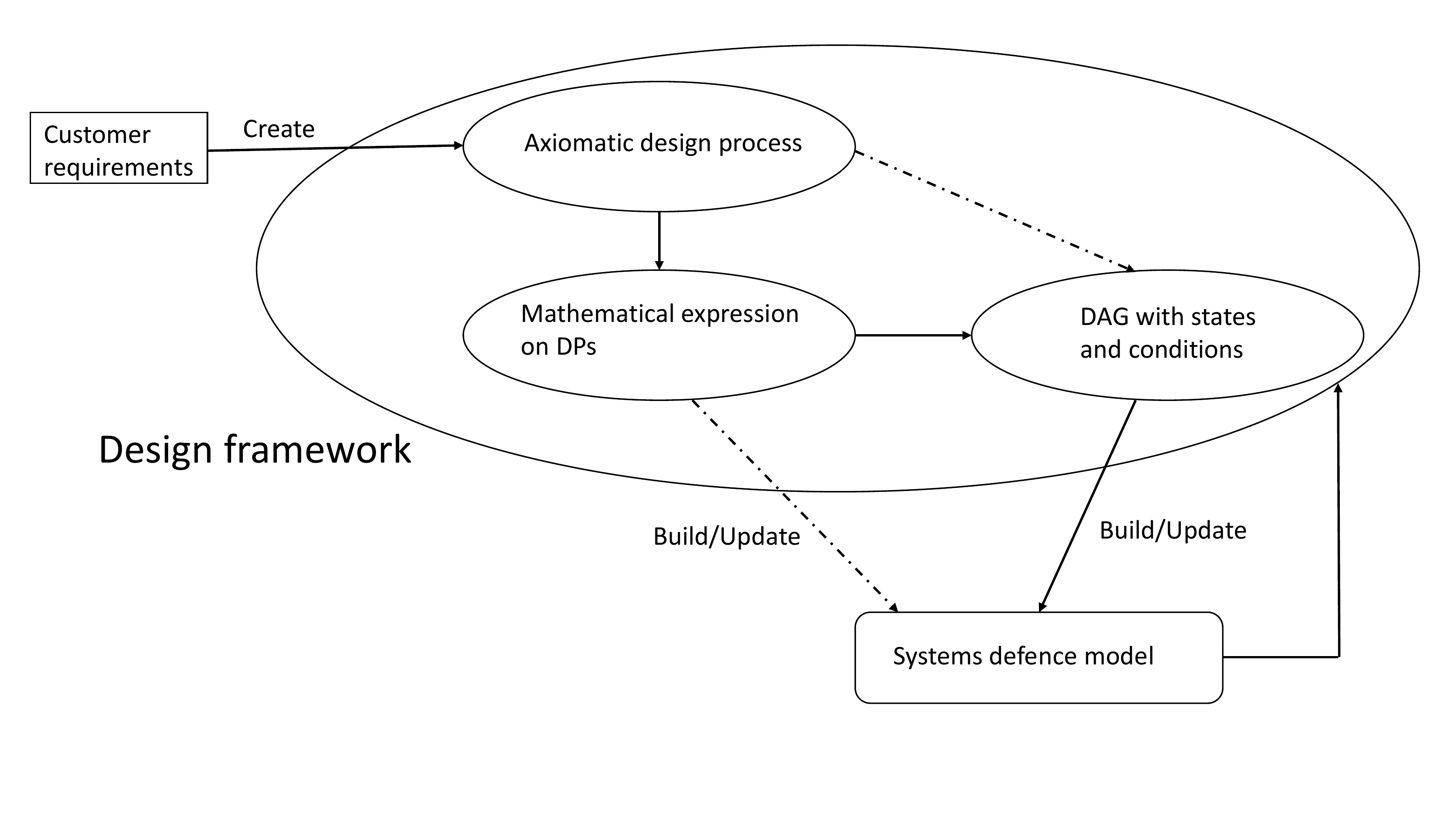}
  \caption{Overview of our proposed method}
  \label{fig:framework_overview}
\end{figure*}

Our proposed systematic design approach consists of three parts: an axiomatic design analysis, the construction of mathematical equations over DPs, and the construction of process graphs, i.e.~directed acyclic graphs~(DAGs) over physical states and conditions. These equations and DAGs characterise invariant relationships between the DPs, and their input/output relations are fed into supervised ML algorithms to learn classifiers that can be used to detect attacks. Figure~\ref{fig:framework_overview} summarises the relationship between these steps.

The workflow begins by transforming the customer's needs (e.g.~build a six-stage water treatment plant) into a list of high-level FRs (e.g.~track water level of tanks) with some associated high-level DPs (e.g.~sensing mechanisms) and PVs (e.g.~value ranges). These are systematically and iteratively decomposed into fine-grained FRs, with particular sensors/actuators as DPs, as guided by the axiomatic design theory. At this stage, the dependencies between particular DPs are transformed into \emph{invariants}, i.e.~mathematical equations over the DP states (e.g.``if DP$_1$ is on and DP$_2$ is low then the system is anomalous''). The design engineers determine which combination of DP states are normal/anomalous, optionally using DAGs with states and conditions to guide their reasoning about paths through the system. Supervised ML algorithms are then trained on the input/output relations of these equations or paths, resulting in classifiers that detect when sensor and actuator data is violating the underlying invariants.

The classifiers, and the invariants underlying them, embody key characteristics of the design as identified by the engineer through a systematic method. We envisage that this has the potential to complement defence mechanisms based on black-box approaches (e.g.~\cite{Chen-Poskitt-Sun18a}), where invariant relationships are based only on observable data after the system has been implemented, and which might not reflect all invariants implied by the design. Furthermore, as our invariants are constructed at the design stage, it may be possible to involve them in early simulations of the processes, and to iteratively modify the system design before it is implemented (Figure~\ref{fig:framework_overview}) if any weak points are identified.

For the rest of this paper, we use the SWaT water purification testbed as a worked example to show how the steps of our approach are applied, and how invariant-based classifiers can be derived.

\subsection{Axiomatic Design Process}

\begin{table*}[!t]
	\centering
	\caption{Top-level decomposition of SWaT}
	\label{tab:higherlevel}
	\begin{tabular}{llll}
		\hline
				No&Functional Requirements (FRs)                  & Design Parameters (DPs)&Process Variables (PVs)          \\
	\hline
		1&	FR1: Feed water to water tanks/systems&	DP1: DOL/VSD Pumps&	Features - Switch (On/Off) and Speed\\
    2&	FR2: Track level of water in tanks	&DP2: Sensing mechanisms&	Value range\\
3&	FR3: Track flow rate of water&	DP3: EMF sensors&	Value range\\
4&	FR4: Monitor chemical properties of water&	DP4: Chemical properties sensors&	Value range\\
5&	FR5: Feed chemicals to water&	DP5: Dosing Pumps&	Switch (On/Off)\\
6&	FR6: Track water pressure&	DP6: Pressure sensors&	Value range\\
7&	FR7: Direct flow of water&	DP7: Motorised valves&	Switch (On/Off)\\
8&	FR8: Track level of chemicals in tanks&	DP8: Level switch&	Value range\\
		\hline
	\end{tabular}
\end{table*}

\begin{table*}[!t]
	\centering
	\caption{Second-level decomposition of FRs and DPs}
	\label{tab:2ndlevel-decomp}
	\resizebox{\textwidth}{!}{
	\begin{tabular}{l l l l}
		\hline
		No&Functional Requirements (FRs) & Design Parameters (DPs)&Process Variables (PVs)          \\
		\hline
		1&	FR1.1: Pump raw water from stage one to UF feed tank in stage three&	DP1.1: P-101,P-102&	On/Off\\
2&	FR1.2: Pump water from stage three to RO feed tank in stage four&	DP1.2: P-301,P-302&	On/Off\\
3&	FR1.3: Pump water from stage four through de-chlorination system&	DP1.3: P-401,P-402&	On/Off\\
4&	FR1.4: Pump (VSD) water from stage five to tanks in stage six	&DP1.4: P-501,P-502	&On/Off\\
5&	FR1.5: Pump water from RO permeate tank to raw water tank in stage one& 	DP1.5: P-601&	On/Off\\
6&	FR1.6: Pump water for UF backwash system&	DP1.6: P-602&	On/Off\\
7&	FR1.7: Pump water for RO/UF cleaning &	DP1.7: P-603&	On/Off\\
8&	FR2.1: Determine water level in raw water tank of stage one&	DP2.1: LIT-101&	0 <= $\alpha$ <= max$_a$\\
9&	FR2.2: Determine water level in UF feed tank of stage three&	DP2.2: LIT-301&	0 <=$\alpha$ <= max$_b$\\
10&	FR2.3: Determine water level in RO feed tank of stage four&	DP2.3: LIT-401&	0 <= $\alpha$ <= max$_c$\\
11&	FR2.4: Determine water level in RO permeate tank of stage six&	DP2.4: LS-601&	Low$_d$ <= $\alpha$ <= High$_d$\\
12&	FR2.5: Determine water level in UF backwash tank of stage six&	DP2.5: LS-602& Low$_e$ <= $\alpha$ <= High$_e$\\
13&	FR2.6: Determine water level in CIP tank of stage six&	DP2.6: LS-603&  Low$_f$ <= $\alpha$ <= High$_f$\\
14&	FR3.1: Measure raw water flow rate in stage one & 	DP3.1: FIT-101	& Low$_g$ <= $\alpha$ <= High$_g$\\
15&	FR3.2: Measure water flow rate in stage two &	DP3.2: FIT-201	&Low$_h$ <= $\alpha$ <= High$_h$\\
16&	FR3.3: Measure water flow rate in stage three&	DP3.3: FIT-301	&Low$_i$ <= $\alpha$ <= High$_i$\\
17&	FR3.4: Measure water flow rate in stage four&	DP3.4: FIT-401	& Low$_j$ <= $\alpha$ <= High$_j$\\
18&	FR3.5: Measure water flow rate in stage five&DP3.5: FIT-501,FIT-502,FIT-503,FIT-504	& Low$_k$ <= $\alpha$ <= High$_k$\\
19&	FR3.6: Measure water flow rate in stage six&	DP3.6: FIT-601	& Low$_l$ <= $\alpha$ <= High$_l$\\
20&	FR4.1: Calculate chemical properties of water&	DP4.1: AIT-201,AIT-202,AIT-203,AIT-301,AIT-302,AIT-303& Low$_m$ <= $\alpha$ <= High$_m$\\
&&AIT-401,AIT-402,AIT-501,AIT-502,AIT-503,AIT-504&\\
21&	FR5.1: Pump chemicals to water&	DP5.1: P-201,P-202,P-203,P-204,P-205,P-206,P-207,P-208,&	On/Off\\
&&P-403,P-404&\\
22&	FR6.1: Measure UF filter differential pressure&	DP6.1: DPIT-301&	0 <= $\alpha$  <= max$_n$\\
23&	FR6.2: Measure RO membrane inlet pressure&	DP6.2: PIT-501&	0 <= $\alpha$  <= max$_o$\\
24&	FR6.3: Measure RO membrane pressure&	DP6.3: PIT-502&	0 <=$\alpha$  <= max$_p$ \\
25&	FR6.4: Measure RO reject pressure&	DP6.4: PIT-503&	0 <= $\alpha$  <= max$_q$\\
26&	FR7.1: Control water flow direction&	DP7.1: MV-101,MV-201,MV-301,MV-302,MV-303,MV-304,&	On/Off\\
&&MV-501,MV-502,MV-503,MV-504&\\
27&	FR8.1: Determine NaCl level in NaCl tank of stage two&	DP8.1: LS-201&	Low$_r$ <= $\alpha$  <= max$_r$\\
28&	FR8.2: Determine HCl level in HCl tank of stage two&	DP8.2: LS-202&	Low$_s$ <= $\alpha$  <= max$_s$\\
29&	FR8.3: Determine NaOCl level in NaOCl tank of stage two&	DP8.3: LS-203&	Low$_t$ <= $\alpha$  <= max$_t$\\ 
30&	FR8.4: Determine NaHSO3 level in NaHSO3 tank of stage four&	DP8.4: LS-401&	Low$_u$ <= $\alpha$  <= max$_u$ \\
		 		\hline
	\end{tabular}}
\end{table*}
\begin{figure}[!t]
       \centering
		\begin{equation}		
		\left(\begin{array}{c} FR1 \\ FR2\\FR3\\FR4\\FR5\\FR6\\FR7\\FR8\end{array}\right) 
		= 
		\left(\begin{array}{cccccccccc} X  &  0& 0 & 0 & 0 &0&  0 &0\\
		 0 &  X & 0 & 0 & 0 & 0 & 0& 0\\ 
		  0  &  0 & X & 0 & 0  & 0 &0&0\\
		   0  &  0 & 0 & X & 0  & 0 & 0&0\\
		   0  &  0 & 0 & 0 & X &  0 & 0&0\\
		    0 &  0 & 0 & 0 & 0 &X& 0 &0\\
		    0 & 0 & 0 & 0 & 0 & 0& X &0 \\
		     0 &  0 & 0 & 0 & 0 &0& 0&X\\\end{array}\right)
		\left(\begin{array}{c} DP1 \\ DP2\\DP3\\DP4\\DP5\\DP6\\DP7\\DP8 \end{array}\right) 
		\label{EQ:equaltion_diagonal}
		\end{equation}
		
	\end{figure}
\begin{figure}[!t]
		\begin{equation}		
		\left(\begin{array}{c} FR1 \\ FR2\\FR3\\FR4\\FR5\\FR6\\FR7\\FR8\end{array}\right) 
		= 
		\left(\begin{array}{cccccccc} X  &  y& y & y & 0 &y&  y &0\\
		 y &  X & y & 0 & 0 & 0 & y& 0\\ 
		  y  &  y & X & 0 & 0  & 0 &y&0\\
		   y  &  0 & 0 & X & y  & 0 & 0&0\\
		   0  &  0 & 0 & y & X &  0 & 0&y\\
		    y &  0 & 0 & 0 & 0 &X& y &0\\
		    y & y & y & 0 & 0 & y& X &0 \\
		     0 &  0 & 0 & 0 & y &0& 0&X\\\end{array}\right)
		\left(\begin{array}{c} DP1 \\ DP2\\DP3\\DP4\\DP5\\DP6\\DP7\\DP8 \end{array}\right) 
		\label{EQ:equaltion_diagonal1}
		\end{equation}
		
\end{figure}
\begin{figure*}[!t]
\begin{equation}		
\resizebox{\textwidth}{!}{		$	\left(\begin{array}{c} FR1.1 \\ FR1.2\\FR1.3\\FR1.4\\FR1.5\\FR1.6\\FR1.7\\FR2.1\\FR2.2\\FR2.3\\FR2.4\\FR2.5\\FR2.6\\FR3.1\\FR3.2\\FR3.3\\FR3.4\\FR3.5\\FR3.6\\FR4.1\\FR5.1\\FR6.1\\FR6.2\\FR6.3\\FR6.4\\FR7.1\\FR8.1\\FR8.2\\FR8.3\\FR8.4\\ \end{array}\right) 
		= 
		\left(\begin{array}{ccccccccccccccccccccccccccccccc}  X  &  0 & 0 & 0 & 0 & 0 & 0 &y& y& 0 & 0 & 0 & 0 &0& y& 0 & 0 & 0 & 0 &y& 0& 0 & 0& 0& 0& y &0 &0 &0 &0   \\
		 0  &  X & 0 & 0 & 0 & 0 & 0 &0& y& y & 0 & 0 & 0 &0& 0& y & 0 & 0 & 0 &y& 0& y & 0& 0& 0& y &0 &0 &0 &0   \\
      		 0  &  0 & X & y & 0 & 0 & 0 &0& 0& y & 0 & 0 & 0 &0& 0& 0 & y & y & 0 &y& 0& 0 & 0& 0& 0& 0 &0 &0 &0 &0   \\  
        		 0  &  0 & y & X & 0 & 0 & 0 &0& 0& 0 & 0 & 0 & 0 &0& 0& 0 & 0 & y & 0 &y& 0& 0 & y& y& y& y &0 &0 &0 &0   \\    
        		         		 0  &  0 & 0 & 0 & X & 0 & 0 &y& 0& 0 & y & 0 & 0 &0& 0& 0 & 0 & 0 & 0 &0& 0& 0 & 0& 0& 0& 0 &0 &0 &0 &0   \\     
                		 0  &  0 & 0 & 0 & 0 & X & 0 &0& 0& 0 & 0 & y & 0 &0& 0& 0 & 0 & 0 & y &0& 0& 0 & 0& 0& 0& y &0 &0 &0 &0   \\  
                 0  &  0 & 0 & 0 & 0 & 0 & X &0& 0& 0 & 0 & 0 & y &0& 0& 0 & 0 & 0 & 0 &0& 0& 0 & 0& 0& 0& 0 &0 &0 &0 &0   \\   
                         y  &  0 & 0 & 0 & y & 0 & 0 &X& 0& 0 & y & 0 & 0 &y& 0& 0 & 0 & 0 & 0 &0& 0& 0 & 0& 0& 0& y &0 &0 &0 &0   \\   
                 y  &  y & 0 & 0 & 0 & 0 & 0 &0& X& 0 & 0 & 0 & 0 &0& y& 0 & 0 & 0 & 0 &0& 0& 0 & 0& 0& 0& y &0 &0 &0 &0   \\   
                 0  &  y & y & 0 & 0 & 0 & 0 &0& 0& X & 0 & 0 & 0 &0& 0& y & y & 0 & 0 &0& 0& 0 & 0& 0& 0& y &0 &0 &0 &0   \\  
                 0  &  0 & 0 & 0 & y & 0 & 0 &y& 0& 0 & X & 0 & 0 &0& 0& 0 & 0 & 0 & 0 &0& 0& 0 & 0& 0& 0& 0 &0 &0 &0 &0   \\  
                 0  &  0 & 0 & 0 & 0 & y & 0 &0& 0& 0 & 0 & X & 0 &0& 0& 0 & 0 & 0 & y &0& 0& 0 & 0& 0& 0& y &0 &0 &0 &0   \\  
                 0  &  0 & 0 & 0 & 0 & 0 & y &0& 0& 0 & 0 & 0 & X &0& 0& 0 & 0 & 0 & 0 &0& 0& 0 & 0& 0& 0& y &0 &0 &0 &0   \\
                 0  &  0 & 0 & 0 & 0 & 0 & 0 &y& 0& 0 & 0 & 0 & 0 &X& 0& 0 & 0 & 0 & 0 &0& 0& 0 & 0& 0& 0& y &0 &0 &0 &0   \\
                 y  &  0 & 0 & 0 & 0 & 0 & 0 &0& y& 0 & 0 & 0 & 0 &0& X& 0 & 0 & 0 & 0 &0& 0& 0 & 0& 0& 0& y &0 &0 &0 &0   \\
                 0  &  y & 0 & 0 & 0 & 0 & 0 &0& 0& y & 0 & 0 & 0 &0& 0& X & 0 & 0 & 0 &0& 0& 0 & 0& 0& 0& y &0 &0 &0 &0   \\   
                 0  &  0 & y & 0 & 0 & 0 & 0 &0& 0& y & 0 & 0 & 0 &0& 0& 0 & X & 0 & 0 &0& 0& 0 & 0& 0& 0& 0 &0 &0 &0 &0   \\   
                 0  &  0 & y & y & 0 & 0 & 0 &0& 0& 0 & 0 & 0 & 0 &0& 0& 0 & 0 & X & 0 &0& 0& 0 & 0& 0& 0& y &0 &0 &0 &0   \\
                 0  &  0 & 0 & 0 & 0 & y & 0 &0& 0& 0 & 0 & y & 0 &0& 0& 0 & 0 & 0 & X &0& 0& 0 & 0& 0& 0& y &0 &0 &0 &0   \\  
                 y  &  y & y & y & 0 & 0 & 0 &0& 0& 0 & 0 & 0 & 0 &0& 0& 0 & 0 & 0 & 0 &X& y& 0 & 0& 0& 0& 0 &0 &0 &0 &0   \\  
                 0  &  0 & 0 & 0 & 0 & 0 & 0 &0& 0& 0 & 0 & 0 & 0 &0& 0& 0 & 0 & 0 & 0 &y& X& 0 & 0& 0& 0& 0 &y &y &y &y   \\  
                 0  &  y & 0 & 0 & 0 & 0 & 0 &0& 0& 0 & 0 & 0 & 0 &0& 0& 0 & 0 & 0 & 0 &0& 0& X & 0& 0& 0& y &0 &0 &0 &0   \\  
                 0  &  0 & 0 & y & 0 & 0 & 0 &0& 0& 0 & 0 & 0 & 0 &0& 0& 0 & 0 & 0 & 0 &0& 0& 0 & X& 0& 0& 0 &0 &0 &0 &0   \\  
                 0  &  0 & 0 & y & 0 & 0 & 0 &0& 0& 0 & 0 & 0 & 0 &0& 0& 0 & 0 & 0 & 0 &0& 0& 0 & 0& X& 0& y &0 &0 &0 &0   \\  
                 0  &  0 & 0 & y & 0 & 0 & 0 &0& 0& 0 & 0 & 0 & 0 &0& 0& 0 & 0 & 0 & 0 &0& 0& 0 & 0& 0& X& y &0 &0 &0 &0   \\ 
                 y  &  y & 0 & y & 0 & y & 0 &y& y& y & 0 & y & y &y& y& y & 0 & y & y &0& 0& y & 0& y& y& X &0 &0 &0 &0   \\  
                 0  &  0 & 0 & 0 & 0 & 0 & 0 &0& 0& 0 & 0 & 0 & 0 &0& 0& 0 & 0 & 0 & 0 &0& y& 0 & 0& 0& 0& 0 &X &0 &0 &0   \\    
                 0  &  0 & 0 & 0 & 0 & 0 & 0 &0& 0& 0 & 0 & 0 & 0 &0& 0& 0 & 0 & 0 & 0 &0& y& 0 & 0& 0& 0& 0 &0 &X &0 &0   \\
                 0  &  0 & 0 & 0 & 0 & 0 & 0 &0& 0& 0 & 0 & 0 & 0 &0& 0& 0 & 0 & 0 & 0 &0& y& 0 & 0& 0& 0& 0 &0 &0 &X &0   \\  
                 0  &  0 & 0 & 0 & 0 & 0 & 0 &0& 0& 0 & 0 & 0 & 0 &0& 0& 0 & 0 & 0 & 0 &0& y& 0 & 0& 0& 0& 0 &0 &0 &0 &X   \\

          \end{array}\right)
		\left(\begin{array}{c} DP1.1\\ DP1.2\\DP1.3\\DP1.4\\DP1.5\\DP1.6\\DP1.7\\DP2.1\\DP2.2\\DP2.3\\DP2.4\\DP2.5\\DP2.6\\DP3.1\\DP3.2\\DP3.3\\DP3.4\\DP3.5\\DP3.6\\DP4.1\\DP5.1\\DP6.1\\DP6.2\\DP6.3\\DP6.4\\DP7.1\\DP8.1\\DP8.2\\DP8.3\\DP8.4 \\\end{array}\right) $}
		\label{EQ:equaltion_SECOND}
		\end{equation}		
	\end{figure*}    

\begin{table*}[!t]
	\centering
	\caption{Linking second-level decomposition of FRs to dependent DPs}
	\label{tab:2ndlevel-decomp-dependentDPs}
	\resizebox{\textwidth}{!}{
	\begin{tabular}{l l l }
		\hline
		No&Functional Requirements (FRs) & Design Parameters (DPs)          \\
		\hline
1	&FR1.1: Pump raw water from stage one to UF feed tank in stage three&	DP1.1: P-101, P-102\\
&&
DP1.1.1: P-101
Other DPs: DP2.1(LIT-101), DP2.2(LIT-301), DP7.1(MV-201)\\
&&
DP1.1.2: P-102
Other DPs: DP2.1(LIT-101), DP2.2(LIT-301), DP7.1(MV-201)\\
2&	FR1.2: Pump water from stage three to RO feed tank in stage four&	DP1.2: P-301, P-302\\
&&DP1.2.1: P-301
Other DPs: DP2.2(LIT-301), DP2.3(LIT-401), DP7.1(MV-302)\\
&&DP1.2.2: P-302
Other DPs: DP2.2(LIT-301), DP2.3(LIT-401), DP7.1(MV-302)\\
3&	FR1.3: Pump water from stage four through de-chlorination system&	DP1.3: P-401, P-402\\
&&DP1.3.1: P-401
Other DPs: DP1.4(P-501,P-502), DP2.3(LIT-401)\\
&&DP1.3.2: P-402
Other DPs: DP1.4(P-501,P-502), DP2.3(LIT-401)\\
4&	FR1.4: Pump (VSD) water from stage five to tanks in stage six&	DP1.4: P-501, P-502\\
&&DP1.4.1: P-501
Other DPs: DP1.3(P-401,P-402), DP7.1(MV-501)\\
&&DP1.4.2: P-502
Other DPs: DP1.3(P-401,P-402), DP7.1(MV-501)\\
5&	FR1.5: Pump water from RO permeate tank to raw water tank in stage one &	DP1.5: P-601
Other DPs: DP2.1(LIT-101), DP2.4(LS-601)\\
6&	FR1.6: Pump water for UF backwash system&	DP1.6: P-602
Other DPs: DP2.5(LS-602), DP7.1(MV-301)\\
7&	FR1.7: Pump water for RO/UF cleaning &	DP1.7: P-603
Other DPs: DP2.6(LS-603)\\
8&	FR2.1: Determine water level in raw water tank of stage one	&DP2.1: LIT-101
Other DPs: DP1.1(P-101,P-102), DP1.5(P-601), DP2.4(LS-601), DP7.1(MV-101)\\
9&	FR2.2: Determine water level in UF feed tank of stage three	&DP2.2: LIT-301
Other DPs: DP1.1(P-101,P-102), DP1.2(P-301,P-302), DP7.1(MV-201)\\
10&	FR2.3: Determine water level in RO feed tank of stage four	&DP2.3: LIT-401
Other DPs: DP1.2(P-301,P-302), DP1.3(P-401,P-402), DP7.1(MV-302)\\
11&	FR2.4: Determine water level in RO permeate tank of stage six&	DP2.4: LS-601
Other DPs: DP1.5(P-601), DP2.1(LIT-101)\\
12&	FR2.5: Determine water level in UF backwash tank of stage six	&DP2.5: LS-602
Other DPs: DP1.6(P-602), DP7.1(MV-301)\\
13&	FR2.6: Determine water level in CIP tank of stage six	&DP2.6: LS-603
Other DPs: DP1.7(P-603), DP7.1(MV-301)\\
14&	FR3.1: Measure raw water flow rate in stage one&  	DP3.1: FIT-101
Other DPs: DP2.1(LIT-101), DP7.1(MV-101)\\
15&	FR3.2: Measure water flow rate in stage two& 	DP3.2: FIT-201
Other DPs: DP1.1(P-101,P-102), DP2.2(LIT-301), DP7.1(MV-201)\\
16	&FR3.3: Measure water flow rate in stage three&	DP3.3: FIT-301
Other DPs: DP1.2(P-301,P-302), DP2.3(LIT-401), DP7.1(MV-302)\\
17&	FR3.4: Measure water flow rate in stage four&	DP3.4: FIT-401
Other DPs: DP1.3(P-401,P-402), DP2.3(LIT-401)\\
18&	FR3.5: Measure water flow rate in stage five&	DP3.5: FIT-501,FIT-502,FIT-503,FIT-504\\
&&DP3.5.1: FIT-501
Other DPs: DP1.3(P-401,P-402)\\
&&DP3.5.2: FIT-502
Other DPs: DP1.4(P-501,P-502)\\
&&DP3.5.3: FIT-503
Other DPs: DP1.4(P-501,P-502)\\
&&DP3.5.4: FIT-504
Other DPs: DP1.3(P-401,P-402)\\
19&	FR3.6: Measure water flow rate in stage six&	DP3.6: FIT-601
Other DPs: DP1.6(P-602), DP2.5(LS-602), DP7.1(MV-301)\\
20&	FR4.1: Calculate chemical properties of water&	DP4.1: AIT-201,AIT-202,AIT-203,AIT-301,AIT-302,AIT-303,\\
&&AIT-401,AIT-402,AIT-501,AIT-502,AIT-503,AIT-504\\
&&DP4.1.1: AIT-201
Other DPs: DP1.1.1(P-101), DP1.1.2(P-102)\\
&&DP4.1.2: AIT-202
Other DPs: DP1.1.1(P-101), DP1.1.2(P-102)\\
&&DP4.1.3: AIT-203
Other DPs: DP1.1.1(P-101), DP1.1.2(P-102)\\
&&DP4.1.4: AIT-301
Other DPs: DP1.2.1(P-301), DP1.2.2(P-302)\\
&&DP4.1.5: AIT-302
Other DPs: DP1.2.1(P-301), DP1.2.2(P-302)\\
&&DP4.1.6: AIT-303
Other DPs: DP1.2.1(P-301), DP1.2.2(P-302)\\
&&DP4.1.7: AIT-401
Other DPs: DP1.3.1(P-401), DP1.3.2(P-402)\\
&&DP4.1.8: AIT-402
Other DPs: DP1.3.1(P-401), DP1.3.2(P-402)\\
&&DP4.1.9: AIT-501
Other DPs: DP1.3.1(P-401), DP1.3.2(P-402)\\
&&DP4.1.10: AIT-502
Other DPs: DP1.3.1(P-401), DP1.3.2(P-402)\\
&&DP4.1.11: AIT-503
Other DPs: DP1.3.1(P-401), DP1.3.2(P-402)\\
&&DP4.1.11: AIT-504
Other DPs: DP1.4.1(P-501), DP1.4.2(P-502)\\
21	&FR5.1: Pump chemicals to water	&DP5.1: P-201,P-202,P-203,P-204,P-205,P-206,P-207,P-208,P-403,P-404\\
&&DP5.1.1: P-201
Other DPs: DP4.1.1(AIT-201), DP7.1.2(MV-201)\\
&&DP5.1.2: P-202
Other DPs: DP4.1.1(AIT-201), DP7.1.2(MV-201)\\
&&DP5.1.3: P-203
Other DPs: DP4.1.2(AIT-202), DP7.1.2(MV-201)\\
&&DP5.1.4: P-204
Other DPs: DP4.1.2(AIT-202), DP7.1.2(MV-201)\\
&&DP5.1.5: P-205
Other DPs: DP4.1.3(AIT-203), DP7.1.2(MV-201)\\
&&DP5.1.6: P-206
Other DPs: DP4.1.3(AIT-203), DP7.1.2(MV-201)\\
&&DP5.1.7: P-207
Other DPs: DP4.1.5(AIT-302)\\
&&DP5.1.8: P-208
Other DPs: DP4.1.5(AIT-302)\\
&&DP5.1.9: P-403
Other DPs: DP4.1.8(AIT-402)\\
&&DP5.1.10: P-404
Other DPs: DP4.1.8(AIT-402)\\
22&	FR6.1: Measure UF filter differential pressure&	DP6.1: DPIT-301
Other DPs: DP1.2.1(P-301), DP1.2.2(P-302), DP7.1(MV-302)\\
23&	FR6.2: Measure RO membrane inlet pressure&	DP6.2: PIT-501
Other DPs: DP1.4.1(P-501), DP1.4.2(P-502)\\
24&	FR6.3: Measure RO membrane pressure&	DP6.3: PIT-502
Other DPs: DP1.4.1(P-501), DP1.4.2(P-502), DP7.1.7(MV-501),\\ &&DP7.1.9(MV-503)\\
25&	FR6.4: Measure RO reject pressure&	DP6.4: PIT-503
Other DPs: DP1.4.1(P-501), DP1.4.2(P-502), DP7.1.8(MV-502),\\ &&DP7.1.10(MV-504)\\

		 		\hline
	\end{tabular}}
\end{table*}

\begin{table*}[!t]
	\centering
	\caption{Linking second-level decomposition of FRs to dependent DPs (continued)}
	\label{tab:2ndlevel-decomp-dependentDPsCont}
	\resizebox{\textwidth}{!}{
	\begin{tabular}{l l l }
		\hline
		No&Functional Requirements (FRs) & Design Parameters (DPs)          \\
		\hline
26&	FR7.1: Control water flow direction&	DP7.1: MV-101,MV-201,MV-301,MV-302,MV-303,MV-304,MV-501,MV-502,MV-503,MV-504\\
&&DP7.1.1: MV-101
Other DPs: DP2.1(LIT-101)\\
&&DP7.1.2: MV-201
Other DPs: DP1.1.1(P-101), DP1.1.2(P-102), DP2.2(LIT-301)\\
&&DP7.1.3: MV-301
Other DPs: DP1.6(P-602), DP2.5(LS-602), DP2.6(LS-603)\\
&&DP7.1.4: MV-302
Other DPs: DP1.2.1(P-301), DP1.2.2(P-302), DP2.3(LIT-401)\\
&&DP7.1.5: MV-303
Other DPs: DP1.2.1(P-301), DP1.2.2(P-302)\\
&&DP7.1.6: MV-304
Other DPs: DP1.2.1(P-301), DP1.2.2(P-302)\\
&&DP7.1.7: MV-501
Other DPs: DP1.4.1(P-501), DP1.4.2(P-502)\\
&&DP7.1.8: MV-502
Other DPs: DP1.4.1(P-501), DP1.4.2(P-502)\\
&&DP7.1.9: MV-503
Other DPs: DP1.4.1(P-501), DP1.4.2(P-502)\\
&&DP7.1.10: MV-504
Other DPs: DP1.4.1(P-501), DP1.4.2(P-502)\\
27	&FR8.1: Determine NaCl level in NaCl tank of stage two	&DP8.1: LS-201
Other DPs: DP5.1.1(P-201), DP5.1.2(P-202)\\
28&	FR8.2: Determine HCl level in HCl tank of stage two&	DP8.2: LS-202
Other DPs: DP5.1.3(P-203), DP5.1.4(P-204)\\
29&	FR8.3: Determine NaOCl level in NaOCl tank of stage two	&DP8.3: LS-203
Other DPs: DP5.1.5(P-205), DP5.1.6(P-206), DP5.1.7(P-207), DP5.1.8(P-208)\\
30&	FR8.4: Determine NaHSO3 level in NaHSO3 tank of stage four	&DP8.4: LS-401
Other DPs: DP5.1.9(P-403), DP5.1.10(P-404)\\

		 		\hline
	\end{tabular}}
\end{table*}

Based on the requirements of SWaT, a top-level decomposition and design are given in Table~\ref{tab:higherlevel}. By the axiomatic design principles, this first level should be a functionally \emph{uncoupled} design guaranteeing that each DP satisfies exactly one FR. This is reflected by the matrix of Equation~(\ref{EQ:equaltion_diagonal}), a diagonal matrix in which each FR is related only to its given DP from Table~\ref{tab:higherlevel}.

Next, the engineer analyses the DPs against the FRs and updates the corresponding Boolean value of the matrix if there is an \emph{information state coupling} between them. This type of coupling differs from the physical coupling used in the conventional axiomatic design theory because it considers state information. Inserting this information state coupling into Equation~(\ref{EQ:equaltion_diagonal}) results in Equation~(\ref{EQ:equaltion_diagonal1}), where $y$ (or $X$ on the diagonal) indicates some dependencies, and a zero ($0$) denotes the absence of them. Note that these dependencies are represented symmetrically: for example, if DP7 is (information state) coupled with FR2, then FR7 is coupled with DP2.

Equation~(\ref{EQ:equaltion_diagonal1}) shows that DP7 is depended on by FR1, FR2, and FR7: for example, a motorised valve (DP7) is opened to feed water into a tank when the level is low (FR2) and the pump is on (FR1), i.e.~the state of some valve depends on the state of some tank and pump. Note however that Equation~(\ref{EQ:equaltion_diagonal1}) presents design information that is very high-level and broadly defined. For instance, FR3---``track flow rate of water''---relates to \emph{multiple} different locations and flow sensors in SWaT. Another example is FR1---``feed water to water tanks/systems''---when in reality there are multiple water pumps in six different stages of SWaT. In order to derive meaningful invariants that relate concrete components of the CPS, our method requires that the top-level design of Equation (\ref{EQ:equaltion_diagonal1}) is iteratively decomposed until there is only a \emph{point-to-point mapping} between each FR and DP. For illustration purposes, such a mapping is shown in the third-level decomposition of Table~\ref{tab:3rdlevel-decomp} at the end of the paper.

For simplicity, rather than use a point-to-point mapping, we decompose the eight FRs of Equation (\ref{EQ:equaltion_diagonal1}) into the 30 FRs of the second-level decomposition in Table~\ref{tab:2ndlevel-decomp}. This is much more concrete than the top-level decomposition as it factors in particular sensors and actuators from different stages, but groups some of them together for convenience (e.g.~P-101 and P-102 are the same DP, as the latter pump is simply a backup for the former). Next, Equation (\ref{EQ:equaltion_SECOND}) is constructed by mapping down the information-state coupling from Equation (\ref{EQ:equaltion_diagonal1}) and adjusting according to the FRs of Table~\ref{tab:2ndlevel-decomp}. At this second level, we use the notational format FR$i$.$j$ and DP$i$.$j$ with $i$ denoting the number from the top-level design and $j$ the number from the second-level.

Finally, Tables~\ref{tab:2ndlevel-decomp-dependentDPs} and ~\ref{tab:2ndlevel-decomp-dependentDPsCont} present the dependent DPs for each second-level FR from Table~\ref{tab:2ndlevel-decomp}, using the information-state coupling as identified by the CPS designer in Equation~(\ref{EQ:equaltion_SECOND}). These sets of dependencies identified in the design can then be used to construct invariants (see Sections~\ref{sec:equations} and \ref{sec:dags}).

Applying axiomatic design here has given the advantage of being able to follow a systematic approach, which guides the analysis of dependencies from a simple top-level design down to one that considers particular sensors and actuators. Furthermore, it allows (in the following) for design-justified invariants to be derived for a complex system like SWaT without the need for any complex mathematical modelling, such as other methods that use Petri-nets \cite{Liu-Zhang-Zhu17a} or Bayesian networks \cite{Hadjsaid-et_al09a} to analyse dependencies.

\subsection{Mathematical Expressions over DPs}\label{sec:equations}

\begin{table}[!t]
  \caption{Mathematical state expressions and assessments}
  \label{tab:freq}
  \begin{tabular}{ll}
    \toprule
    State Expression (Input) & Assessment (Output)\\
    \midrule
    !MV-101 and !LIT-101&	Anomaly\\
    !MV-101 and LIT-101&	No anomaly\\
    MV-101 and !LIT-101&	No anomaly\\
    MV-101 and LIT-101&	Anomaly\\

  \bottomrule
\end{tabular}
\end{table}

With the sets of dependencies between DPs in Tables~\ref{tab:2ndlevel-decomp-dependentDPs} and ~\ref{tab:2ndlevel-decomp-dependentDPsCont}, we can then construct mathematical state expressions that characterise the \emph{invariant} relationships between them. Consider, for example, row 26 of Table~\ref{tab:2ndlevel-decomp-dependentDPsCont}: here, DP7.1.1 expresses that motorised valve MV-101 has a dependency on LIT-101 (and vice versa).

Given these two components, we then construct mathematical equations to characterise their invariant relationship, i.e.~the combinations of states they will always be in during normal operation. These consist of the states that the DPs are in (`inputs') and assignments of anomalous or non-anomalous (`output'). In the case of LIT-101, we use the low/high thresholds to determine two discrete states of interest, then relate them against the possible discrete values of MV-101 (open or closed). These equations are given in Table~\ref{tab:freq}, where MV-101 (resp.~!MV-101) denotes that the valve is open (resp.~closed), and LIT-101 (resp.~!LIT-101) denotes that the tank level is high (resp.~low). The table also reflects the judgement of an engineer as to which of these four combinations reflect anomalous configurations. For example, if the valve is open and the tank level is high, this is anomalous as it could cause the tank to overflow.

At this stage of our work, when constructing equations, we assume that all DPs have one of two states, and thus the total number of equations to analyse will be $2^n$ where $n$ is the number of components involved. For large numbers of $n$, the number of equations may grow too large to manually complete, hence our use of decision trees and other supervised ML algorithms in Section~\ref{sec:study} to generalise from the inputs/outputs that we do have.

\subsection{DAGs with States and Conditions}\label{sec:dags}
We also use the directed acyclic graph (DAG) concept of graph theory~\cite{Bondy-Murty08a} for sketching \emph{process graphs} to assist the engineer in analysing the relationship between states of DPs. In particular, we use nodes to represent the states of DPs and edges to represent the conditions that cause the states of DPs to change.

\begin{figure*}[!t]
  \includegraphics[trim=0 140 0 0,clip,width = 0.9\textwidth ]{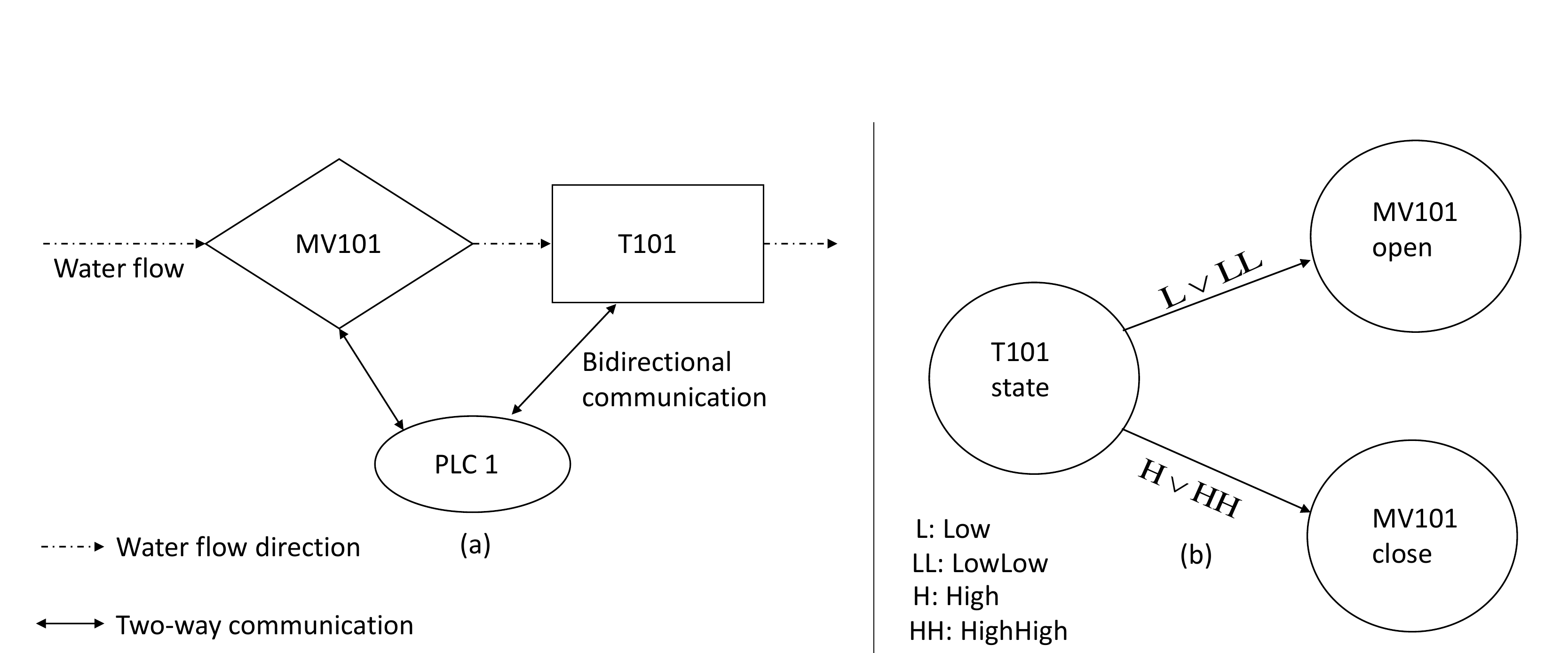}
  \caption{(a) Subsystem in stage one of SWaT;  (b) A process graph in stage one of SWaT for normal cases}
  \label{fig:process_graphs}
\end{figure*}

\begin{figure}[t]
 \centering
  \includegraphics[trim=80 160 0 0,clip,width = 0.9\textwidth ]{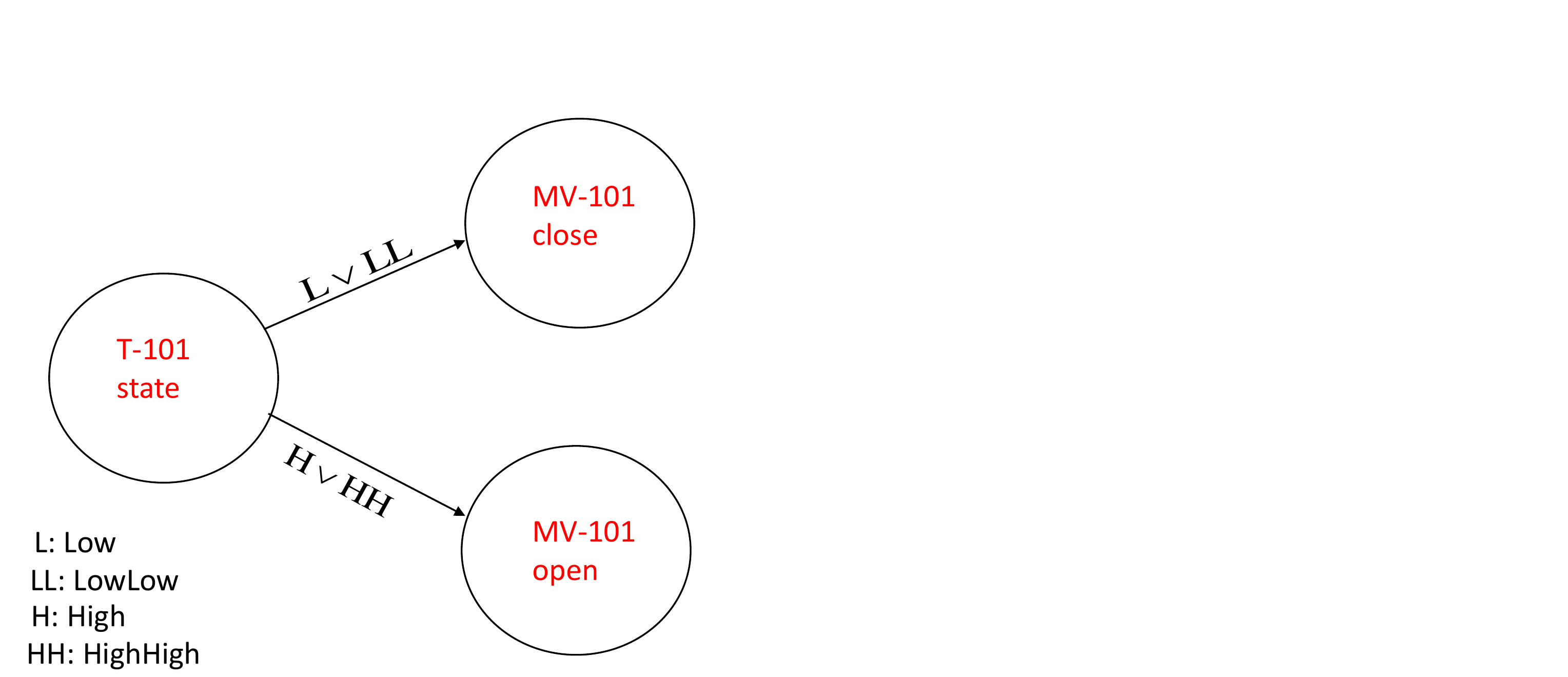}
  \caption{A process graph depicting anomalous cases}
  \label{fig:process_graphs_anomalies}
\end{figure}

Figure~\ref{fig:process_graphs}(a) shows a subsystem in stage one of SWaT and the interconnection of some dependent devices identified from the axiomatic design analysis (row 26 of Table~\ref{tab:2ndlevel-decomp-dependentDPsCont}), along with the physical tank (T-101) and PLC that controls the components. Figures~\ref{fig:process_graphs}(b) and \ref{fig:process_graphs_anomalies} then visualise how the state of one DP (MV-101) changes according to the other (LIT-101), for both the normal and the (inverted) anomalous cases. In both graphs, we consider the cases when the tank level is below one of its low thresholds (Low or LowLow), or above one of its high thresholds (High or HighHigh). MV-101 can be either open or closed, and the graphs reflect the states it should switch to according to different level thresholds of the tank. The idea is that the designer traverses through the different paths in the normal case of Figures~\ref{fig:process_graphs}(b), and then concludes that all other possible paths are abnormal cases, thus deriving Figure~\ref{fig:process_graphs_anomalies}. For example, if the tank level is low (L or LL), then the correct behaviour is for the valve to open and allow water to flow in; remaining closed (risking an underflow scenario) is thus anomalous.

These paths in Figures~\ref{fig:process_graphs}(b) and \ref{fig:process_graphs_anomalies} provide direct and indirect information for deciding the outcome of an attack, and the identified anomaly cases should be investigated further to prevent successful attacks. This information is useful because it gives the designer the possibility to employ different types of defence mechanisms at certain locations based on cost and impact analysis. The graphs provide visual representations of paths which may be easier to communicate and analyse in comparison to mathematical expressions. A deeper analysis of Figures~\ref{fig:process_graphs}(b) and \ref{fig:process_graphs_anomalies} by the designer may lead to the discovery of a successful concrete attack such as spoofing of data. For example: if LIT-101 for T-101 is spoofed to be at H instead of L, then P-101 starts pumping water out of T-101. In a short time, T-101 is going to be emptied. A prolonged period of pumping water from an empty tank can damage the pumps. A solution to prevent such an attack is to use data encryption for network communication between devices. As this weakness is determined at the design phase of the system, the cost of protecting against such an attack is likely to be less than if the attack had been discovered after the implementation of hardware and software.

\section{Preliminary Evaluation}\label{sec:study}

In this section, we present a preliminary study in which we assess how effective our invariants are for implementing \emph{attack detection mechanisms} in SWaT. First, we derive two invariants from our design framework: one concerning sensors and actuators in stage one only, and another concerning sensors and actuators across three stages (1--3). Next, we train supervised machine learning algorithms (e.g.~decision trees, na\"{i}ve Bayes) on the inputs/outputs of the equations and process graphs associated with the invariants. The training results in \emph{classifiers} that can judge actual sensor/actuator data as `normal' or `anomalous' with respect to the invariant relationships they were trained on. Finally, we deploy these classifiers on both the SWaT dataset~\cite{CPS-Datasets} as well as the actual testbed to assess how effective they are at detecting some relevant attacks.

\substepseparator

\noindent\textbf{Experiment and Results.} The first invariant (\#1) we derived concerns the relationship between MV-101 and LIT-101. This dependency was identified from our design analysis, in particular, from DP7.1.1, row 26 of Table~\ref{tab:2ndlevel-decomp-dependentDPsCont}. The completed mathematical equations and their corresponding process graphs are given in Table~\ref{tab:freq}, Figure~\ref{fig:process_graphs}(b), and Figure~\ref{fig:process_graphs_anomalies}. As this dependency involved only two components, it was feasible to completely classify all Boolean combinations of their discretised values as normal or anomalous. This might not be the case for larger sets of dependencies, which is why supervised learning algorithms may be required in general.

The second invariant (\#2) concerns the relationship between LIT-301, MV-201, P-101, and P-102. These dependencies were also identified from our design analysis, in particular, from DP7.1.2, row 26 of Table~\ref{tab:2ndlevel-decomp-dependentDPsCont}. Mathematical equations and process graphs were constructed for these components in an analogous way to the first invariant.

We trained both supervised decision tree and na\"{i}ve Bayes learning algorithms on the inputs/outputs of the mathematical equations and process graphs of these invariants. We applied the classifiers to the SWaT dataset~\cite{CPS-Datasets} and found that \emph{no false positives} were reported, i.e.~at no point did `normal' data in the labelled dataset get misclassified as `anomalous'. Likewise, when implemented for the actual testbed, we found that the classifiers did not report any false positives throughout our experiments. We investigated the effectiveness of the classifiers against relevant attacks in both the SWaT dataset as well as on the testbed itself. We found that our decision tree classifiers achieved $100\%$ accuracy in the prediction of normal/anomalous labels throughout these attacks, and that our na\"{i}ve Bayes classifiers achieved $100\%$ and $93.75\%$ accuracy for the two invariants respectively. The two attack tests we considered for the SWaT testbed (of potentially many more) are discussed below.

While we are motivated by the results, we are actively working on building a larger set of invariants/classifiers to be sure that the results do generalise, and are keen to compare the effectiveness of our invariants against those derived using other approaches, e.g.~those that used only data and not the system design at all.

\substepseparator

\noindent\textbf{Attack Test \#1.} The following attack scenario was successfully detected using the SWaT testbed.

\emph{Components to attack:} Motorised valve MV-101, and level indicator transmitter LIT-101 (both in stage one).

\emph{Objective:} Damage or reduce the reliability of MV-101.

\emph{Launch state:} At time $t$, the water level of LIT-101 is above 800mm.

\emph{Attack:} At time $t\!+\!+$, an attacker begins \emph{manually} turning MV-101 on and off several times.

\emph{Results:} Successfully detected. The attack violated the classifier of invariant \#1 for MV-101, LIT-101.

SWaT's defences could be strengthened against this attack by overriding attempts to manually turn on MV-101 if the water level reported by LIT-101 is above its high (H) or critically high (HH) thresholds. This would block the attack, although would rely on the assumption that the value of LIT-101 is correct can be trusted (additional invariants and classifiers concerning LIT-101 could help to mitigate this threat).

\substepseparator

\noindent\textbf{Attack Test \#2.} This second attack test case was also successfully executed and detected using the SWaT testbed.

\emph{Components to attack:} Pumps P-101, P-102, and motorised valve MV-201.

\emph{Objective:} Damage or reduce the reliability of P-101, P-102, and MV-201.

\emph{Launch state:} At time $t$, the water level reported by LIT-301 is above 1000mm.

\emph{Attack:} At time $t\!+\!+$, an attacker manually turns on MV-201.  Next, the attacker manually starts pump P-102. Finally, the attacker manually starts pump P-101. 

\emph{Results:} Successfully detected. The attack violated the classifier of invariant \#2 for LIT-301, MV-201, P-101, P-102.

SWaT's defences could be strengthened by overriding attempts to turn on MV-201 if the water level reported by LIT-301 is above its high (H) or critically high (HH) thresholds. This would prevent the attack from happening, but again, would require additional mechanisms or improvements to the design to ensure that the value of LIT-301 can be trusted.

\section{Conclusion}\label{sec:conclusion}
In this position paper, we proposed a method for systematically deriving invariants from the functional requirements of a CPS, in order to: (1)~integrate security concerns at the design stage; (2)~ensure that invariants can be traced from requirements through to implemented defence mechanisms; (3)~identify all invariants implicit in the design, not just those represented in datasets; and (4)~potentially save costs by identifying weak points before the CPS is built. Inspired by the axiomatic design methodology for systems, our method iteratively analyses dependencies in a given CPS to construct equations and process graphs that represent invariant relations between sensor readings and actuator states. We demonstrated the potential use of this approach in a preliminary study on the SWaT water treatment testbed, implementing checkers for two invariants using decision trees, and finding that they detected relevant attacks without false positives.

Our method provides a step-by-step approach from requirements to the implementation of invariant checkers, and does not require any complex mathematical modelling or dataset-based training. This may help to ensure that detected attacks or system malfunctions can be \emph{explained}, not just in terms of the equations or process graph that they violated, but also in terms of the functional requirements that they were derived from, providing systems engineers with a rich amount of context.

In ongoing work, we plan extend our preliminary study for SWaT by implementing more invariant-based checkers from our requirements analysis, and evaluating them against a wider array of attacks. We also plan to compare the effectiveness of our invariants against those derived by other approaches (both automated and manual ones) in order to better understand the added security benefits of analysing the CPS design directly. Finally, we plan to assess the generalisability of our approach by applying it to other industrial control systems (e.g.~the WADI water distribution plant), and potentially to CPSs from other domains, such as building management or healthcare systems.

\begin{acks}
We are grateful to the anonymous referees of CPSS'20 for their helpful and encouraging comments. This research / project is supported by the National Research Foundation, Singapore, under its National Satellite of Excellence Programme ``Design Science and Technology for Secure Critical Infrastructure'' (Award Number: NSoE\_DeST-SCI2019-0004). Any opinions, findings and conclusions or recommendations expressed in this material are those of the author(s) and do not reflect the views of National Research Foundation, Singapore.
\end{acks}

\bibliographystyle{ACM-Reference-Format}
\balance
\bibliography{references}


\begin{thebibliography}{50}


\ifx \showCODEN    \undefined \def \showCODEN     #1{\unskip}     \fi
\ifx \showDOI      \undefined \def \showDOI       #1{#1}\fi
\ifx \showISBNx    \undefined \def \showISBNx     #1{\unskip}     \fi
\ifx \showISBNxiii \undefined \def \showISBNxiii  #1{\unskip}     \fi
\ifx \showISSN     \undefined \def \showISSN      #1{\unskip}     \fi
\ifx \showLCCN     \undefined \def \showLCCN      #1{\unskip}     \fi
\ifx \shownote     \undefined \def \shownote      #1{#1}          \fi
\ifx \showarticletitle \undefined \def \showarticletitle #1{#1}   \fi
\ifx \showURL      \undefined \def \showURL       {\relax}        \fi
\providecommand\bibfield[2]{#2}
\providecommand\bibinfo[2]{#2}
\providecommand\natexlab[1]{#1}
\providecommand\showeprint[2][]{arXiv:#2}

\bibitem[\protect\citeauthoryear{??}{CPS}{2020}]%
        {CPS-Datasets}
 \bibinfo{year}{2020}\natexlab{}.
\newblock \bibinfo{title}{{iTrust Labs: Datasets}}.
\newblock
  \bibinfo{howpublished}{{\url{https://itrust.sutd.edu.sg/itrust-labs_datasets/}}}.
\newblock
\newblock
\shownote{Accessed: July 2020.}


\bibitem[\protect\citeauthoryear{??}{SWa}{2020}]%
        {SWaT-Reference}
 \bibinfo{year}{2020}\natexlab{}.
\newblock \bibinfo{title}{{Secure Water Treatment (SWaT)}}.
\newblock
  \bibinfo{howpublished}{{\url{https://itrust.sutd.edu.sg/itrust-labs-home/itrust-labs_swat/}}}.
\newblock
\newblock
\shownote{Accessed: July 2020.}


\bibitem[\protect\citeauthoryear{Adepu, Brasser, Garcia, Rodler, Davi, Sadeghi,
  and Zonouz}{Adepu et~al\mbox{.}}{2020}]%
        {Adepu-et_al20a}
\bibfield{author}{\bibinfo{person}{Sridhar Adepu}, \bibinfo{person}{Ferdinand
  Brasser}, \bibinfo{person}{Luis Garcia}, \bibinfo{person}{Michael Rodler},
  \bibinfo{person}{Lucas Davi}, \bibinfo{person}{Ahmad{-}Reza Sadeghi}, {and}
  \bibinfo{person}{Saman~A. Zonouz}.} \bibinfo{year}{2020}\natexlab{}.
\newblock \showarticletitle{Control Behavior Integrity for Distributed
  Cyber-Physical Systems}. In \bibinfo{booktitle}{\emph{Proc.\ {ACM/IEEE}
  International Conference on Cyber-Physical Systems ({ICCPS} 2020)}}.
  \bibinfo{publisher}{{IEEE}}, \bibinfo{pages}{30--40}.
\newblock


\bibitem[\protect\citeauthoryear{Adepu and Mathur}{Adepu and Mathur}{2016a}]%
        {Adepu-Mathur16b}
\bibfield{author}{\bibinfo{person}{Sridhar Adepu} {and} \bibinfo{person}{Aditya
  Mathur}.} \bibinfo{year}{2016}\natexlab{a}.
\newblock \showarticletitle{Distributed Detection of Single-Stage Multipoint
  Cyber Attacks in a Water Treatment Plant}. In
  \bibinfo{booktitle}{\emph{Proc.\ {ACM} Asia Conference on Computer and
  Communications Security (AsiaCCS 2016)}}. \bibinfo{publisher}{{ACM}},
  \bibinfo{pages}{449--460}.
\newblock


\bibitem[\protect\citeauthoryear{Adepu and Mathur}{Adepu and Mathur}{2016b}]%
        {Adepu-Mathur16a}
\bibfield{author}{\bibinfo{person}{Sridhar Adepu} {and} \bibinfo{person}{Aditya
  Mathur}.} \bibinfo{year}{2016}\natexlab{b}.
\newblock \showarticletitle{Using Process Invariants to Detect Cyber Attacks on
  a Water Treatment System}. In \bibinfo{booktitle}{\emph{Proc.\ International
  Conference on ICT Systems Security and Privacy Protection (SEC 2016)}}
  \emph{(\bibinfo{series}{IFIP AICT})}, Vol.~\bibinfo{volume}{471}.
  \bibinfo{publisher}{Springer}, \bibinfo{pages}{91--104}.
\newblock


\bibitem[\protect\citeauthoryear{Adepu and Mathur}{Adepu and Mathur}{2018}]%
        {Adepu-Mathur18b}
\bibfield{author}{\bibinfo{person}{Sridhar Adepu} {and} \bibinfo{person}{Aditya
  Mathur}.} \bibinfo{year}{2018}\natexlab{}.
\newblock \showarticletitle{Distributed Attack Detection in a Water Treatment
  Plant: Method and Case Study}.
\newblock \bibinfo{journal}{\emph{IEEE Transactions on Dependable and Secure
  Computing}} (\bibinfo{year}{2018}).
\newblock


\bibitem[\protect\citeauthoryear{Aggarwal, Karimibiuki, Pattabiraman, and
  Ivanov}{Aggarwal et~al\mbox{.}}{2018}]%
        {Aggarwal-et_al18a}
\bibfield{author}{\bibinfo{person}{Ekta Aggarwal}, \bibinfo{person}{Mehdi
  Karimibiuki}, \bibinfo{person}{Karthik Pattabiraman}, {and}
  \bibinfo{person}{Andr{\'{e}} Ivanov}.} \bibinfo{year}{2018}\natexlab{}.
\newblock \showarticletitle{{CORGIDS:} {A} Correlation-based Generic Intrusion
  Detection System}. In \bibinfo{booktitle}{\emph{Proc.\ Workshop on
  Cyber-Physical Systems Security and PrivaCy (CPS-SPC 2018)}}.
  \bibinfo{publisher}{{ACM}}, \bibinfo{pages}{24--35}.
\newblock


\bibitem[\protect\citeauthoryear{Ahmed, Ochoa, Zhou, Mathur, Qadeer, Murguia,
  and Ruths}{Ahmed et~al\mbox{.}}{2018a}]%
        {Ahmed-et_al18a}
\bibfield{author}{\bibinfo{person}{Chuadhry~Mujeeb Ahmed},
  \bibinfo{person}{Mart{\'{\i}}n Ochoa}, \bibinfo{person}{Jianying Zhou},
  \bibinfo{person}{Aditya~P. Mathur}, \bibinfo{person}{Rizwan Qadeer},
  \bibinfo{person}{Carlos Murguia}, {and} \bibinfo{person}{Justin Ruths}.}
  \bibinfo{year}{2018}\natexlab{a}.
\newblock \showarticletitle{\emph{NoisePrint}: Attack Detection Using Sensor
  and Process Noise Fingerprint in Cyber Physical Systems}. In
  \bibinfo{booktitle}{\emph{Proc.\ Asia Conference on Computer and
  Communications Security (AsiaCCS 2018)}}. \bibinfo{publisher}{{ACM}},
  \bibinfo{pages}{483--497}.
\newblock


\bibitem[\protect\citeauthoryear{Ahmed, Zhou, and Mathur}{Ahmed
  et~al\mbox{.}}{2018b}]%
        {Ahmed-et_al18b}
\bibfield{author}{\bibinfo{person}{Chuadhry~Mujeeb Ahmed},
  \bibinfo{person}{Jianying Zhou}, {and} \bibinfo{person}{Aditya~P. Mathur}.}
  \bibinfo{year}{2018}\natexlab{b}.
\newblock \showarticletitle{Noise Matters: Using Sensor and Process Noise
  Fingerprint to Detect Stealthy Cyber Attacks and Authenticate sensors in
  {CPS}}. In \bibinfo{booktitle}{\emph{Proc.\ Annual Computer Security
  Applications Conference (ACSAC 2018)}}. \bibinfo{publisher}{{ACM}},
  \bibinfo{pages}{566--581}.
\newblock


\bibitem[\protect\citeauthoryear{Aoudi, Iturbe, and Almgren}{Aoudi
  et~al\mbox{.}}{2018}]%
        {Aoudi-et_al18a}
\bibfield{author}{\bibinfo{person}{Wissam Aoudi}, \bibinfo{person}{Mikel
  Iturbe}, {and} \bibinfo{person}{Magnus Almgren}.}
  \bibinfo{year}{2018}\natexlab{}.
\newblock \showarticletitle{Truth Will Out: Departure-Based Process-Level
  Detection of Stealthy Attacks on Control Systems}. In
  \bibinfo{booktitle}{\emph{Proc.\ {ACM} {SIGSAC} Conference on Computer and
  Communications Security (CCS 2018)}}. \bibinfo{publisher}{{ACM}},
  \bibinfo{pages}{817--831}.
\newblock


\bibitem[\protect\citeauthoryear{Bondy and Murty}{Bondy and Murty}{2008}]%
        {Bondy-Murty08a}
\bibfield{author}{\bibinfo{person}{J.~Adrian Bondy} {and}
  \bibinfo{person}{Uppaluri S.~R. Murty}.} \bibinfo{year}{2008}\natexlab{}.
\newblock \bibinfo{booktitle}{\emph{Graph Theory}}.
\newblock \bibinfo{publisher}{Springer}.
\newblock


\bibitem[\protect\citeauthoryear{C{\'{a}}rdenas, Amin, Lin, Huang, Huang, and
  Sastry}{C{\'{a}}rdenas et~al\mbox{.}}{2011}]%
        {Cardenas-et_al11a}
\bibfield{author}{\bibinfo{person}{Alvaro~A. C{\'{a}}rdenas},
  \bibinfo{person}{Saurabh Amin}, \bibinfo{person}{Zong{-}Syun Lin},
  \bibinfo{person}{Yu{-}Lun Huang}, \bibinfo{person}{Chi{-}Yen Huang}, {and}
  \bibinfo{person}{Shankar Sastry}.} \bibinfo{year}{2011}\natexlab{}.
\newblock \showarticletitle{Attacks against process control systems: risk
  assessment, detection, and response}. In \bibinfo{booktitle}{\emph{Proc.\
  {ACM} Asia Conference on Computer and Communications Security (AsiaCCS
  2011)}}. \bibinfo{publisher}{{ACM}}, \bibinfo{pages}{355--366}.
\newblock


\bibitem[\protect\citeauthoryear{Carrasco and Wu}{Carrasco and Wu}{2019}]%
        {Carrasco-Wu19a}
\bibfield{author}{\bibinfo{person}{Mayra Alexandra~Macas Carrasco} {and}
  \bibinfo{person}{Chunming Wu}.} \bibinfo{year}{2019}\natexlab{}.
\newblock \showarticletitle{An Unsupervised Framework for Anomaly Detection in
  a Water Treatment System}. In \bibinfo{booktitle}{\emph{Proc.\ {IEEE}
  International Conference On Machine Learning And Applications ({ICMLA}
  2019)}}. \bibinfo{publisher}{{IEEE}}, \bibinfo{pages}{1298--1305}.
\newblock


\bibitem[\protect\citeauthoryear{Chen, Poskitt, and Sun}{Chen
  et~al\mbox{.}}{2016}]%
        {Chen-Poskitt-Sun16a}
\bibfield{author}{\bibinfo{person}{Yuqi Chen}, \bibinfo{person}{Christopher~M.
  Poskitt}, {and} \bibinfo{person}{Jun Sun}.} \bibinfo{year}{2016}\natexlab{}.
\newblock \showarticletitle{Towards Learning and Verifying Invariants of
  Cyber-Physical Systems by Code Mutation}. In \bibinfo{booktitle}{\emph{Proc.\
  International Symposium on Formal Methods (FM 2016)}}
  \emph{(\bibinfo{series}{LNCS})}, Vol.~\bibinfo{volume}{9995}.
  \bibinfo{publisher}{Springer}, \bibinfo{pages}{155--163}.
\newblock


\bibitem[\protect\citeauthoryear{Chen, Poskitt, and Sun}{Chen
  et~al\mbox{.}}{2018}]%
        {Chen-Poskitt-Sun18a}
\bibfield{author}{\bibinfo{person}{Yuqi Chen}, \bibinfo{person}{Christopher~M.
  Poskitt}, {and} \bibinfo{person}{Jun Sun}.} \bibinfo{year}{2018}\natexlab{}.
\newblock \showarticletitle{Learning from Mutants: Using Code Mutation to Learn
  and Monitor Invariants of a Cyber-Physical System}. In
  \bibinfo{booktitle}{\emph{Proc.\ {IEEE} Symposium on Security and Privacy
  (S{\&}P 2018)}}. \bibinfo{publisher}{{IEEE} Computer Society},
  \bibinfo{pages}{648--660}.
\newblock


\bibitem[\protect\citeauthoryear{Chen, Poskitt, Sun, Adepu, and Zhang}{Chen
  et~al\mbox{.}}{2019}]%
        {Chen-Poskitt-et_al19a}
\bibfield{author}{\bibinfo{person}{Yuqi Chen}, \bibinfo{person}{Christopher~M.
  Poskitt}, \bibinfo{person}{Jun Sun}, \bibinfo{person}{Sridhar Adepu}, {and}
  \bibinfo{person}{Fan Zhang}.} \bibinfo{year}{2019}\natexlab{}.
\newblock \showarticletitle{Learning-Guided Network Fuzzing for Testing
  Cyber-Physical System Defences}. In \bibinfo{booktitle}{\emph{Proc.\ IEEE/ACM
  International Conference on Automated Software Engineering (ASE 2019)}}.
  \bibinfo{publisher}{{IEEE} Computer Society}, \bibinfo{pages}{962--973}.
\newblock


\bibitem[\protect\citeauthoryear{Chen, Xuan, Poskitt, Sun, and Zhang}{Chen
  et~al\mbox{.}}{2020}]%
        {Chen-Xuan-Poskitt-et_al20a}
\bibfield{author}{\bibinfo{person}{Yuqi Chen}, \bibinfo{person}{Bohan Xuan},
  \bibinfo{person}{Christopher~M. Poskitt}, \bibinfo{person}{Jun Sun}, {and}
  \bibinfo{person}{Fan Zhang}.} \bibinfo{year}{2020}\natexlab{}.
\newblock \showarticletitle{Active Fuzzing for Testing and Securing
  Cyber-Physical Systems}. In \bibinfo{booktitle}{\emph{Proc.\ ACM SIGSOFT
  International Symposium on Software Testing and Analysis (ISSTA 2020)}}.
  \bibinfo{publisher}{ACM}.
\newblock


\bibitem[\protect\citeauthoryear{Cheng, Tian, and Yao}{Cheng
  et~al\mbox{.}}{2017}]%
        {Cheng-Tian-Yao17a}
\bibfield{author}{\bibinfo{person}{Long Cheng}, \bibinfo{person}{Ke Tian},
  {and} \bibinfo{person}{Danfeng~(Daphne) Yao}.}
  \bibinfo{year}{2017}\natexlab{}.
\newblock \showarticletitle{{Orpheus}: Enforcing Cyber-Physical Execution
  Semantics to Defend Against Data-Oriented Attacks}. In
  \bibinfo{booktitle}{\emph{Proc.\ Annual Computer Security Applications
  Conference (ACSAC 2017)}}. \bibinfo{publisher}{{ACM}},
  \bibinfo{pages}{315--326}.
\newblock


\bibitem[\protect\citeauthoryear{Choi, Lee, Aafer, Fei, Tu, Zhang, Xu, and
  Xinyan}{Choi et~al\mbox{.}}{2018}]%
        {Choi-et_al18a}
\bibfield{author}{\bibinfo{person}{Hongjun Choi}, \bibinfo{person}{Wen{-}Chuan
  Lee}, \bibinfo{person}{Yousra Aafer}, \bibinfo{person}{Fan Fei},
  \bibinfo{person}{Zhan Tu}, \bibinfo{person}{Xiangyu Zhang},
  \bibinfo{person}{Dongyan Xu}, {and} \bibinfo{person}{Xinyan Xinyan}.}
  \bibinfo{year}{2018}\natexlab{}.
\newblock \showarticletitle{Detecting Attacks Against Robotic Vehicles: {A}
  Control Invariant Approach}. In \bibinfo{booktitle}{\emph{Proc.\ {ACM}
  {SIGSAC} Conference on Computer and Communications Security ({CCS} 2018)}}.
  \bibinfo{publisher}{{ACM}}, \bibinfo{pages}{801--816}.
\newblock


\bibitem[\protect\citeauthoryear{Das, Adepu, and Zhou}{Das
  et~al\mbox{.}}{2020}]%
        {Das-Adepu-Zhou20a}
\bibfield{author}{\bibinfo{person}{Tanmoy~Kanti Das}, \bibinfo{person}{Sridhar
  Adepu}, {and} \bibinfo{person}{Jianying Zhou}.}
  \bibinfo{year}{2020}\natexlab{}.
\newblock \showarticletitle{Anomaly detection in Industrial Control Systems
  using Logical Analysis of Data}.
\newblock \bibinfo{journal}{\emph{Computers \& Security}}  \bibinfo{volume}{96}
  (\bibinfo{year}{2020}).
\newblock


\bibitem[\protect\citeauthoryear{Feng, Palleti, Mathur, and Chana}{Feng
  et~al\mbox{.}}{2019}]%
        {Feng-et_al19a}
\bibfield{author}{\bibinfo{person}{Cheng Feng}, \bibinfo{person}{Venkata~Reddy
  Palleti}, \bibinfo{person}{Aditya Mathur}, {and} \bibinfo{person}{Deeph
  Chana}.} \bibinfo{year}{2019}\natexlab{}.
\newblock \showarticletitle{A Systematic Framework to Generate Invariants for
  Anomaly Detection in Industrial Control Systems}. In
  \bibinfo{booktitle}{\emph{Proc.\ Annual Network and Distributed System
  Security Symposium (NDSS 2019)}}. \bibinfo{publisher}{The Internet Society}.
\newblock


\bibitem[\protect\citeauthoryear{Formby, Srinivasan, Leonard, Rogers, and
  Beyah}{Formby et~al\mbox{.}}{2016}]%
        {Formby-et_al16a}
\bibfield{author}{\bibinfo{person}{David Formby}, \bibinfo{person}{Preethi
  Srinivasan}, \bibinfo{person}{Andrew~M. Leonard},
  \bibinfo{person}{Jonathan~D. Rogers}, {and} \bibinfo{person}{Raheem~A.
  Beyah}.} \bibinfo{year}{2016}\natexlab{}.
\newblock \showarticletitle{Who's in Control of Your Control System? Device
  Fingerprinting for Cyber-Physical Systems}. In
  \bibinfo{booktitle}{\emph{Proc.\ Annual Network and Distributed System
  Security Symposium ({NDSS} 2016)}}. \bibinfo{publisher}{The Internet
  Society}.
\newblock


\bibitem[\protect\citeauthoryear{Giraldo, Urbina, Cardenas, Valente, Faisal,
  Ruths, Tippenhauer, Sandberg, and Candell}{Giraldo et~al\mbox{.}}{2018}]%
        {Giraldo-et_al18a}
\bibfield{author}{\bibinfo{person}{Jairo Giraldo}, \bibinfo{person}{David~I.
  Urbina}, \bibinfo{person}{Alvaro Cardenas}, \bibinfo{person}{Junia Valente},
  \bibinfo{person}{Mustafa~Amir Faisal}, \bibinfo{person}{Justin Ruths},
  \bibinfo{person}{Nils~Ole Tippenhauer}, \bibinfo{person}{Henrik Sandberg},
  {and} \bibinfo{person}{Richard Candell}.} \bibinfo{year}{2018}\natexlab{}.
\newblock \showarticletitle{A Survey of Physics-Based Attack Detection in
  Cyber-Physical Systems}.
\newblock \bibinfo{journal}{\emph{Comput. Surveys}} \bibinfo{volume}{51},
  \bibinfo{number}{4} (\bibinfo{year}{2018}), \bibinfo{pages}{76:1--76:36}.
\newblock


\bibitem[\protect\citeauthoryear{Goh, Adepu, Junejo, and Mathur}{Goh
  et~al\mbox{.}}{2016}]%
        {Goh-et_al16a}
\bibfield{author}{\bibinfo{person}{Jonathan Goh}, \bibinfo{person}{Sridhar
  Adepu}, \bibinfo{person}{Khurum~Nazir Junejo}, {and} \bibinfo{person}{Aditya
  Mathur}.} \bibinfo{year}{2016}\natexlab{}.
\newblock \showarticletitle{A Dataset to Support Research in the Design of
  Secure Water Treatment Systems}. In \bibinfo{booktitle}{\emph{Proc.\
  International Conference on Critical Information Infrastructures Security
  (CRITIS 2016)}}.
\newblock


\bibitem[\protect\citeauthoryear{Goh, Adepu, Tan, and Lee}{Goh
  et~al\mbox{.}}{2017}]%
        {Goh_et-al17a}
\bibfield{author}{\bibinfo{person}{Jonathan Goh}, \bibinfo{person}{Sridhar
  Adepu}, \bibinfo{person}{Marcus Tan}, {and} \bibinfo{person}{Zi~Shan Lee}.}
  \bibinfo{year}{2017}\natexlab{}.
\newblock \showarticletitle{Anomaly detection in cyber physical systems using
  recurrent neural networks}. In \bibinfo{booktitle}{\emph{Proc.\ International
  Symposium on High Assurance Systems Engineering (HASE 2017)}}.
  \bibinfo{publisher}{{IEEE}}, \bibinfo{pages}{140--145}.
\newblock


\bibitem[\protect\citeauthoryear{Gu, Formby, Ji, Cam, and Beyah}{Gu
  et~al\mbox{.}}{2018}]%
        {Gu-et_al18a}
\bibfield{author}{\bibinfo{person}{Qinchen Gu}, \bibinfo{person}{David Formby},
  \bibinfo{person}{Shouling Ji}, \bibinfo{person}{Hasan Cam}, {and}
  \bibinfo{person}{Raheem~A. Beyah}.} \bibinfo{year}{2018}\natexlab{}.
\newblock \showarticletitle{Fingerprinting for Cyber-Physical System Security:
  Device Physics Matters Too}.
\newblock \bibinfo{journal}{\emph{{IEEE} Security {\&} Privacy}}
  \bibinfo{volume}{16}, \bibinfo{number}{5} (\bibinfo{year}{2018}),
  \bibinfo{pages}{49--59}.
\newblock


\bibitem[\protect\citeauthoryear{Hadjsaid, Tranchita, Rozel, Viziteu, and
  Caire}{Hadjsaid et~al\mbox{.}}{2009}]%
        {Hadjsaid-et_al09a}
\bibfield{author}{\bibinfo{person}{Nouredine Hadjsaid},
  \bibinfo{person}{Carolina Tranchita}, \bibinfo{person}{Beno{\^{\i}}t Rozel},
  \bibinfo{person}{Maria Viziteu}, {and} \bibinfo{person}{Raphael Caire}.}
  \bibinfo{year}{2009}\natexlab{}.
\newblock \showarticletitle{Modeling cyber and physical interdependencies -
  Application in ICT and power grids}. In \bibinfo{booktitle}{\emph{Proc.\
  IEEE/PES Power Systems Conference and Exposition (PSCE 2009)}}.
  \bibinfo{publisher}{IEEE}.
\newblock


\bibitem[\protect\citeauthoryear{Harada, Yamagata, Mizuno, and Choi}{Harada
  et~al\mbox{.}}{2017}]%
        {Harada-et_al17a}
\bibfield{author}{\bibinfo{person}{Yoshiyuki Harada}, \bibinfo{person}{Yoriyuki
  Yamagata}, \bibinfo{person}{Osamu Mizuno}, {and} \bibinfo{person}{Eun{-}Hye
  Choi}.} \bibinfo{year}{2017}\natexlab{}.
\newblock \showarticletitle{Log-Based Anomaly Detection of {CPS} Using a
  Statistical Method}. In \bibinfo{booktitle}{\emph{Proc.\ International
  Workshop on Empirical Software Engineering in Practice (IWESEP 2017)}}.
  \bibinfo{publisher}{{IEEE}}, \bibinfo{pages}{1--6}.
\newblock


\bibitem[\protect\citeauthoryear{Hassanzadeh, Rasekh, Galelli, Aghashahi,
  Taormina, Ostfeld, and Banks}{Hassanzadeh et~al\mbox{.}}{2019}]%
        {Hassanzadeh-et_al19a}
\bibfield{author}{\bibinfo{person}{Amin Hassanzadeh}, \bibinfo{person}{Amin
  Rasekh}, \bibinfo{person}{Stefano Galelli}, \bibinfo{person}{Mohsen
  Aghashahi}, \bibinfo{person}{Riccardo Taormina}, \bibinfo{person}{Avi
  Ostfeld}, {and} \bibinfo{person}{M.~Katherine Banks}.}
  \bibinfo{year}{2019}\natexlab{}.
\newblock \showarticletitle{A Review of Cybersecurity Incidents in the Water
  Sector}.
\newblock \bibinfo{journal}{\emph{Journal of Environmental Engineering}}
  (\bibinfo{date}{09} \bibinfo{year}{2019}).
\newblock


\bibitem[\protect\citeauthoryear{He, Raghavan, Hu, Chai, and Lee}{He
  et~al\mbox{.}}{2019}]%
        {He-et_al19a}
\bibfield{author}{\bibinfo{person}{Zecheng He}, \bibinfo{person}{Aswin
  Raghavan}, \bibinfo{person}{Guangyuan Hu}, \bibinfo{person}{Sek~M. Chai},
  {and} \bibinfo{person}{Ruby~B. Lee}.} \bibinfo{year}{2019}\natexlab{}.
\newblock \showarticletitle{Power-Grid Controller Anomaly Detection with
  Enhanced Temporal Deep Learning}. In \bibinfo{booktitle}{\emph{Proc.\ {IEEE}
  International Conference On Trust, Security And Privacy In Computing And
  Communications (TrustCom 2019)}}. \bibinfo{publisher}{{IEEE}},
  \bibinfo{pages}{160--167}.
\newblock


\bibitem[\protect\citeauthoryear{{ICS-CERT Alert}}{{ICS-CERT Alert}}{2016}]%
        {ICS-Cert-Alert16a}
\bibfield{author}{\bibinfo{person}{{ICS-CERT Alert}}.}
  \bibinfo{year}{2016}\natexlab{}.
\newblock \bibinfo{title}{Cyber-Attack Against {Ukrainian} Critical
  Infrastructure}.
\newblock
  \bibinfo{howpublished}{\url{https://ics-cert.us-cert.gov/alerts/IR-ALERT-H-16-056-01}}.
\newblock
\newblock
\shownote{document number: {IR-ALERT-H-16-056-01}.}


\bibitem[\protect\citeauthoryear{Inoue, Yamagata, Chen, Poskitt, and Sun}{Inoue
  et~al\mbox{.}}{2017}]%
        {Inoue-et_al17a}
\bibfield{author}{\bibinfo{person}{Jun Inoue}, \bibinfo{person}{Yoriyuki
  Yamagata}, \bibinfo{person}{Yuqi Chen}, \bibinfo{person}{Christopher~M.
  Poskitt}, {and} \bibinfo{person}{Jun Sun}.} \bibinfo{year}{2017}\natexlab{}.
\newblock \showarticletitle{Anomaly Detection for a Water Treatment System
  Using Unsupervised Machine Learning}. In \bibinfo{booktitle}{\emph{Proc.\
  {IEEE} International Conference on Data Mining Workshops (ICDMW 2017): Data
  Mining for Cyberphysical and Industrial Systems (DMCIS 2017)}}.
  \bibinfo{publisher}{IEEE}, \bibinfo{pages}{1058--1065}.
\newblock


\bibitem[\protect\citeauthoryear{Kandjani, Tavana, Bernus, Wen, and
  Mohtarami}{Kandjani et~al\mbox{.}}{2015}]%
        {Kandjani-et_al15a}
\bibfield{author}{\bibinfo{person}{Hadi Kandjani}, \bibinfo{person}{Madjid
  Tavana}, \bibinfo{person}{Peter Bernus}, \bibinfo{person}{Lian Wen}, {and}
  \bibinfo{person}{Amir Mohtarami}.} \bibinfo{year}{2015}\natexlab{}.
\newblock \showarticletitle{Using extended Axiomatic Design theory to reduce
  complexities in Global Software Development projects}.
\newblock \bibinfo{journal}{\emph{Computers in Industry}}  \bibinfo{volume}{67}
  (\bibinfo{year}{2015}), \bibinfo{pages}{86--96}.
\newblock


\bibitem[\protect\citeauthoryear{Kim, Yun, and Kim}{Kim et~al\mbox{.}}{2019}]%
        {Kim-Yun-Kim19a}
\bibfield{author}{\bibinfo{person}{Jonguk Kim}, \bibinfo{person}{Jeong{-}Han
  Yun}, {and} \bibinfo{person}{Hyoung~Chun Kim}.}
  \bibinfo{year}{2019}\natexlab{}.
\newblock \showarticletitle{Anomaly Detection for Industrial Control Systems
  Using Sequence-to-Sequence Neural Networks}. In
  \bibinfo{booktitle}{\emph{Proc.\ International Workshop on the Security of
  Industrial Control Systems and Cyber-Physical Systems (CyberICPS 2019)}}
  \emph{(\bibinfo{series}{LNCS})}, Vol.~\bibinfo{volume}{11980}.
  \bibinfo{publisher}{Springer}, \bibinfo{pages}{3--18}.
\newblock


\bibitem[\protect\citeauthoryear{Kneib and Huth}{Kneib and Huth}{2018}]%
        {Kneib-Huth18a}
\bibfield{author}{\bibinfo{person}{Marcel Kneib} {and}
  \bibinfo{person}{Christopher Huth}.} \bibinfo{year}{2018}\natexlab{}.
\newblock \showarticletitle{Scission: Signal Characteristic-Based Sender
  Identification and Intrusion Detection in Automotive Networks}. In
  \bibinfo{booktitle}{\emph{Proc.\ {ACM} {SIGSAC} Conference on Computer and
  Communications Security (CCS 2018)}}. \bibinfo{publisher}{{ACM}},
  \bibinfo{pages}{787--800}.
\newblock


\bibitem[\protect\citeauthoryear{Kravchik and Shabtai}{Kravchik and
  Shabtai}{2018}]%
        {Kravchik-Shabtai18a}
\bibfield{author}{\bibinfo{person}{Moshe Kravchik} {and} \bibinfo{person}{Asaf
  Shabtai}.} \bibinfo{year}{2018}\natexlab{}.
\newblock \showarticletitle{Detecting Cyber Attacks in Industrial Control
  Systems Using Convolutional Neural Networks}. In
  \bibinfo{booktitle}{\emph{Proc.\ Workshop on Cyber-Physical Systems Security
  and PrivaCy (CPS-SPC 2018)}}. \bibinfo{publisher}{{ACM}},
  \bibinfo{pages}{72--83}.
\newblock


\bibitem[\protect\citeauthoryear{Leyden}{Leyden}{2016}]%
        {Leyden16a}
\bibfield{author}{\bibinfo{person}{John Leyden}.}
  \bibinfo{year}{2016}\natexlab{}.
\newblock \showarticletitle{Water treatment plant hacked, chemical mix changed
  for tap supplies}.
\newblock \bibinfo{journal}{\emph{The Register}} (\bibinfo{year}{2016}).
\newblock
\urldef\tempurl%
\url{https://www.theregister.com/2016/03/24/water_utility_hacked/}
\showURL{%
\tempurl}
\newblock
\shownote{Accessed: July\ 2020.}


\bibitem[\protect\citeauthoryear{Lin, Adepu, Verwer, and Mathur}{Lin
  et~al\mbox{.}}{2018}]%
        {Lin-et_al18a}
\bibfield{author}{\bibinfo{person}{Qin Lin}, \bibinfo{person}{Sridhar Adepu},
  \bibinfo{person}{Sicco Verwer}, {and} \bibinfo{person}{Aditya Mathur}.}
  \bibinfo{year}{2018}\natexlab{}.
\newblock \showarticletitle{{TABOR:} {A} Graphical Model-based Approach for
  Anomaly Detection in Industrial Control Systems}. In
  \bibinfo{booktitle}{\emph{Proc.\ Asia Conference on Computer and
  Communications Security (AsiaCCS 2018)}}. \bibinfo{publisher}{{ACM}},
  \bibinfo{pages}{525--536}.
\newblock


\bibitem[\protect\citeauthoryear{Liu, Zhang, and Zhu}{Liu
  et~al\mbox{.}}{2017}]%
        {Liu-Zhang-Zhu17a}
\bibfield{author}{\bibinfo{person}{Xiaoxue Liu}, \bibinfo{person}{Jiexin
  Zhang}, {and} \bibinfo{person}{Peidong Zhu}.}
  \bibinfo{year}{2017}\natexlab{}.
\newblock \showarticletitle{Modeling cyber-physical attacks based on
  probabilistic colored Petri nets and mixed-strategy game theory}.
\newblock \bibinfo{journal}{\emph{International Journal of Critical
  Infrastructure Protection}}  \bibinfo{volume}{16} (\bibinfo{year}{2017}),
  \bibinfo{pages}{13--25}.
\newblock


\bibitem[\protect\citeauthoryear{Mathur and Tippenhauer}{Mathur and
  Tippenhauer}{2016}]%
        {Mathur-Tippenhauer16a}
\bibfield{author}{\bibinfo{person}{Aditya~P. Mathur} {and}
  \bibinfo{person}{Nils~Ole Tippenhauer}.} \bibinfo{year}{2016}\natexlab{}.
\newblock \showarticletitle{SWaT: a water treatment testbed for research and
  training on {ICS} security}. In \bibinfo{booktitle}{\emph{Proc.\
  International Workshop on Cyber-physical Systems for Smart Water Networks
  (CySWater@CPSWeek 2016)}}. \bibinfo{publisher}{{IEEE} Computer Society},
  \bibinfo{pages}{31--36}.
\newblock


\bibitem[\protect\citeauthoryear{Matt}{Matt}{2012}]%
        {Matt12a}
\bibfield{author}{\bibinfo{person}{Dominik~T. Matt}.}
  \bibinfo{year}{2012}\natexlab{}.
\newblock \showarticletitle{Application of Axiomatic Design principles to
  control complexity dynamics in a mixed-model assembly system: a case
  analysis}.
\newblock \bibinfo{journal}{\emph{International Journal of Production
  Research}}  \bibinfo{volume}{50}, Article \bibinfo{articleno}{7}
  (\bibinfo{year}{2012}), \bibinfo{numpages}{12}~pages.
\newblock


\bibitem[\protect\citeauthoryear{Mohsen and Cekecek}{Mohsen and
  Cekecek}{2000}]%
        {Mohsen-Cekecek00a}
\bibfield{author}{\bibinfo{person}{Hassan~A. Mohsen} {and}
  \bibinfo{person}{Ethem Cekecek}.} \bibinfo{year}{2000}\natexlab{}.
\newblock \showarticletitle{Thoughts on the use of axiomatic designs within the
  product development process}. In \bibinfo{booktitle}{\emph{Proc.\
  International Conference on Axiomatic Design (ICAD 2000)}}.
\newblock


\bibitem[\protect\citeauthoryear{Narayanan and Bobba}{Narayanan and
  Bobba}{2018}]%
        {Narayanan-Bobba18a}
\bibfield{author}{\bibinfo{person}{Vedanth Narayanan} {and}
  \bibinfo{person}{Rakesh~B. Bobba}.} \bibinfo{year}{2018}\natexlab{}.
\newblock \showarticletitle{Learning Based Anomaly Detection for Industrial Arm
  Applications}. In \bibinfo{booktitle}{\emph{Proc.\ Workshop on Cyber-Physical
  Systems Security and PrivaCy (CPS-SPC 2018)}}. \bibinfo{publisher}{{ACM}},
  \bibinfo{pages}{13--23}.
\newblock


\bibitem[\protect\citeauthoryear{Palleti, Joseph, and Silva}{Palleti
  et~al\mbox{.}}{2018}]%
        {Palleti-Joseph-Silva18a}
\bibfield{author}{\bibinfo{person}{Venkata~Reddy Palleti},
  \bibinfo{person}{Jude~Victor Joseph}, {and} \bibinfo{person}{Arlindo Silva}.}
  \bibinfo{year}{2018}\natexlab{}.
\newblock \showarticletitle{A contribution of axiomatic design principles to
  the analysis and impact of attacks on critical infrastructures}.
\newblock \bibinfo{journal}{\emph{International Journal of Critical
  Infrastructure Protection}}  \bibinfo{volume}{23} (\bibinfo{year}{2018}),
  \bibinfo{pages}{21--32}.
\newblock


\bibitem[\protect\citeauthoryear{Pasqualetti, Dorfler, and Bullo}{Pasqualetti
  et~al\mbox{.}}{2011}]%
        {Pasqualetti-Dorfler-Bullo11a}
\bibfield{author}{\bibinfo{person}{Fabio Pasqualetti}, \bibinfo{person}{Florian
  Dorfler}, {and} \bibinfo{person}{Francesco Bullo}.}
  \bibinfo{year}{2011}\natexlab{}.
\newblock \showarticletitle{{Cyber-physical attacks in power networks: Models,
  fundamental limitations and monitor design}}. In
  \bibinfo{booktitle}{\emph{Proc.\ {IEEE} Conference on Decision and Control
  and European Control Conference (CDC-ECC 2011)}}.
  \bibinfo{publisher}{{IEEE}}, \bibinfo{pages}{2195--2201}.
\newblock


\bibitem[\protect\citeauthoryear{Schneider and B{\"{o}}ttinger}{Schneider and
  B{\"{o}}ttinger}{2018}]%
        {Schneider-Boettinger18a}
\bibfield{author}{\bibinfo{person}{Peter Schneider} {and}
  \bibinfo{person}{Konstantin B{\"{o}}ttinger}.}
  \bibinfo{year}{2018}\natexlab{}.
\newblock \showarticletitle{High-Performance Unsupervised Anomaly Detection for
  Cyber-Physical System Networks}. In \bibinfo{booktitle}{\emph{Proc.\ Workshop
  on Cyber-Physical Systems Security and PrivaCy (CPS-SPC 2018)}}.
  \bibinfo{publisher}{{ACM}}, \bibinfo{pages}{1--12}.
\newblock


\bibitem[\protect\citeauthoryear{Suh}{Suh}{2001}]%
        {Suh01a}
\bibfield{author}{\bibinfo{person}{Nam~Pyo Suh}.}
  \bibinfo{year}{2001}\natexlab{}.
\newblock \bibinfo{booktitle}{\emph{Axiomatic Design: Advances and
  Applications}}.
\newblock \bibinfo{publisher}{Oxford University Press}.
\newblock


\bibitem[\protect\citeauthoryear{Wijaya, Aniche, and Mathur}{Wijaya
  et~al\mbox{.}}{2020}]%
        {Wijaya-Aniche-Mathur20a}
\bibfield{author}{\bibinfo{person}{Herman Wijaya},
  \bibinfo{person}{Maur{\'{\i}}cio Aniche}, {and} \bibinfo{person}{Aditya
  Mathur}.} \bibinfo{year}{2020}\natexlab{}.
\newblock \showarticletitle{Domain-Based Fuzzing for Supervised Learning of
  Anomaly Detection in Cyber-Physical Systems}. In
  \bibinfo{booktitle}{\emph{Proc.\ International Workshop on Engineering and
  Cybersecurity of Critical Systems (EnCyCriS 2020)}}.
  \bibinfo{publisher}{{ACM}}.
\newblock


\bibitem[\protect\citeauthoryear{Yang, Li, Lin, Chen, and Sun}{Yang
  et~al\mbox{.}}{2020}]%
        {Yang-et_al20a}
\bibfield{author}{\bibinfo{person}{Kai Yang}, \bibinfo{person}{Qiang Li},
  \bibinfo{person}{Xiaodong Lin}, \bibinfo{person}{Xin Chen}, {and}
  \bibinfo{person}{Limin Sun}.} \bibinfo{year}{2020}\natexlab{}.
\newblock \showarticletitle{iFinger: Intrusion Detection in Industrial Control
  Systems via Register-Based Fingerprinting}.
\newblock \bibinfo{journal}{\emph{IEEE Journal of Selected Areas in
  Communications}} \bibinfo{volume}{38}, \bibinfo{number}{5}
  (\bibinfo{year}{2020}), \bibinfo{pages}{955--967}.
\newblock


\bibitem[\protect\citeauthoryear{Zhu, Hu, Koren, and Marin}{Zhu
  et~al\mbox{.}}{2008}]%
        {Zhu-et_al08a}
\bibfield{author}{\bibinfo{person}{Xiaowei Zhu}, \bibinfo{person}{S.~Jack Hu},
  \bibinfo{person}{Yoram Koren}, {and} \bibinfo{person}{Samuel~P. Marin}.}
  \bibinfo{year}{2008}\natexlab{}.
\newblock \showarticletitle{Modeling of manufacturing complexity in mixed-model
  assembly lines}.
\newblock \bibinfo{journal}{\emph{Journal of Manufacturing Science and
  Engineering}}  \bibinfo{volume}{130}, Article \bibinfo{articleno}{5}
  (\bibinfo{year}{2008}), \bibinfo{numpages}{10}~pages.
\newblock


\end{thebibliography}

\appendix

\begin{table*}[!t]
	\centering
	\caption{Third-level decomposition of FRs and DPs}
	\label{tab:3rdlevel-decomp}
	\resizebox{\textwidth}{!}{ 
	\begin{tabular}{l l l l}
		\hline
		No&Functional Requirements (FRs) & Design Parameters (DPs)&Process Variables (PVs)          \\
		\hline
1&	FR1.1.1: Feed raw water from stage one to UF feed tank in stage three using pump P-101&	DP1.1.1: P-101&	On/Off\\
2&	FR1.1.2: Feed raw water from stage one to UF feed tank in stage three using pump P-102&	DP1.1.2: P-102&	On/Off\\
3&	FR1.2.1: Feed water from stage three to RO feed tank in stage four using pump P-301&	DP1.2.1: P-301&	On/Off\\
4&	FR1.2.2: Feed water from stage three to RO feed tank in stage four using pump P-302&	DP1.2.2: P-302&	On/Off\\
5&	FR1.3.1: Pump water from stage four through de-chlorination system using pump P-401&	DP1.3.1: P-401&	On/Off\\
6&	FR1.3.2: Pump water from stage four through de-chlorination system using pump P-402&	DP1.3.2: P-402&	On/Off\\
7&	FR1.4.1: Pump (VSD) water from stage five to tanks in stage six using pump P-501&	DP1.4.1: P-501&	On/Off\\
8&	FR1.4.2: Pump (VSD) water from stage five to tanks in stage six using pump P-502&	DP1.4.2: P-502&	On/Off\\
9&	FR3.5.1: Compute RO membrane inlet flow meter in stage five&	DP3.5.1: FIT-501&	0 <= $\alpha$ <= max$_{k1}$\\
10&	FR3.5.2: Compute RO permeate flow meter in stage five&	DP3.5.2: FIT-502&	0 <= $\alpha$ <= max$_{k2}$\\
11&	FR3.5.3: Compute RO reject flow meter in stage five&	DP3.5.3: FIT-503&	0 <= $\alpha$ <= max$_{k3}$\\
12&	FR3.5.4: Compute RO re-circulation flow meter in stage five&	DP3.5.4: FIT-504&	0 <= $\alpha$ <= max$_{k4}$\\
13&	FR4.1.1: Calculate chemical dosing conductivity of water in stage two&	DP4.1.1: AIT-201&	0 <= $\alpha$ <= max$_{m1}$\\
14&	FR4.1.2: Calculate chemical dosing pH of water in stage two	&DP4.1.2: AIT-202&	0 <= $\alpha$ <= max$_{m2}$\\
15&	FR4.1.3: Calculate chemical dosing ORP of water in stage two	&DP4.1.3: AIT-203&	0 <= $\alpha$ <= max$_{m3}$\\
16&	FR4.1.4: Calculate UF permeate pH of water in stage three	&DP4.1.4: AIT-301&	0 <= $\alpha$ <= max$_{m4}$\\
17&	FR4.1.5: Calculate UF permeate ORP of water in stage three	&DP4.1.5: AIT-302&	0 <= $\alpha$ <= max$_{m5}$\\
18&	FR4.1.6: Calculate UF permeate conductivity of water in stage three&	DP4.1.6: AIT-303&	0 <= $\alpha$ <= max$_{m6}$\\
19&	FR4.1.7: Calculate RO feed hardness of water in stage four	&DP4.1.7: AIT-401&	0 <= $\alpha$ <= max$_{m7}$\\
20&	FR4.1.8: Calculate RO ORP of water in stage four&	DP4.1.8: AIT-402&	0 <= $\alpha$ <= max$_{m8}$\\
21&	FR4.1.9: Calculate RO feed pH of water in stage five	&DP4.1.9: AIT-501&	0 <= $\alpha$ <= max$_{m9}$\\
22&	FR4.1.10: Calculate RO feed ORP of water in stage five	&DP4.1.10: AIT-502&	0 <= $\alpha$ <= max$_{m10}$\\
23&	FR4.1.11: Calculate RO feed conductivity of water in stage five&	DP4.1.11: AIT-503&	0 <= $\alpha$ <= max$_{m11}$\\
24&	FR4.1.12: Calculate RO permeate conductivity of water in stage five&	DP4.1.12: AIT-504&	0 <= $\alpha$ <= max$_{m12}$\\
25&	FR5.1.1: Feed NaCl dosing in stage two using pump P-201	&DP5.1.1: P-201&	On/Off\\
26&	FR5.1.2: Feed NaCl dosing in stage two using pump P-202	&DP5.1.2: P-202&	On/Off\\
27&	FR5.1.3: Feed HCl dosing in stage two using pump P-203	&DP5.1.3: P-203&	On/Off\\
28&	FR5.1.4: Feed HCl dosing in stage two using pump P-204	&DP5.1.4: P-204&	On/Off\\
29&	FR5.1.5: Feed NaOCl dosing in stage two using pump P-205 (FAC)	&DP5.1.5: P-205&	On/Off\\
30&	FR5.1.6: Feed NaOCl dosing in stage two using pump P-206 (FAC)	&DP5.1.6: P-206&	On/Off\\
31&	FR5.1.7: Feed NaOCl dosing to stage three UF cleaning using pump P-207 (UF)&	DP5.1.7: P-207&	On/Off\\
32&	FR5.1.8: Feed NaOCl dosing to stage three UF cleaning using pump P-208 (UF)&	DP5.1.8: P-208&	On/Off\\
33&	FR5.1.9: Feed NaHSO3 dosing in stage four using pump P-403 	&DP5.1.9: P-403&	On/Off\\
34&	FR5.1.10: Feed NaHSO3 dosing in stage four using pump P-404 	&DP5.1.10: P-404&	On/Off\\
35&	FR7.1.1: Direct raw water inlet in stage one&	DP7.1.1: MV-101&	On/Off\\
36&	FR7.1.2: Direct water flow in stage two&	DP7.1.2: MV-201	&On/Off\\
37&	FR7.1.3: Direct UF backwash in stage three&	DP7.1.3: MV-301	&On/Off\\
38&	FR7.1.4: Direct UF feed water in stage three&	DP7.1.4: MV-302&	On/Off\\
39&	FR7.1.5: Direct UF backwash drain in stage three&	DP7.1.5: MV-303&	On/Off\\
40&	FR7.1.6: Direct UF drain in stage three&	DP7.1.6: MV-304	&On/Off\\
41&	FR7.1.7: Direct RO permeate in stage five&	DP7.1.7: MV-501	&On/Off\\
42&	FR7.1.8: Direct RO backwash in stage five&	DP7.1.8: MV-502	&On/Off\\
43&	FR7.1.9: Direct RO permeate reject in stage five&	DP7.1.9: MV-503&	On/Off\\
44&	FR7.1.10: Direct RO reject in stage five&	DP7.1.10: MV-504	&On/Off\\

		 		\hline
	\end{tabular} }
\end{table*}

\end{document}